\documentclass[floats,floatfix,showpacs,amssymb,prd,twocolumn,superscriptaddress,nofootinbib,nolongbibliography,reprint]{revtex4-2}

\usepackage{amssymb,amsmath,verbatim,mathtools,needspace,enumitem,etoolbox,graphicx,physics,microtype,afterpage}
\usepackage{gensymb}

\usepackage[dvipsnames, usenames]{xcolor}

\definecolor{linkcolor}{rgb}{0.0,0.3,0.5}
\usepackage[unicode, colorlinks=true, linkcolor=linkcolor, citecolor=linkcolor, filecolor=linkcolor,urlcolor=linkcolor, pdfusetitle]{hyperref}
\usepackage{textcomp}
\usepackage[all]{hypcap}
\usepackage[T1]{fontenc}
\usepackage[utf8]{inputenc}
\usepackage{tabularx}
\usepackage[normalem]{ulem}
\usepackage{placeins}
\usepackage{float}
\usepackage{verbatim}
\usepackage{graphicx}
\usepackage {tikz}
\usetikzlibrary {positioning}
\usepackage{multirow}
\definecolor{processblue}{cmyk}{0.96,0,0,0}

\renewcommand{\vec}[1]{\mathbf{#1}}

\def\aj{\rm{Astron.~J.}}

\def\apj{\rm{Astrophys.~J.}}

\def\aap{\rm{Astron.~Astrophys.}}
\def\aapr{\rm{A\&A~Rev.}}

\def\mnras{\rm{Mon.~Not.~R.~Astron.~Soc.}}
\def\prd{\rm{Phys.~Rev.~D}}

\def\pasj{\rm{PASJ}}

\newcommand{\dallas}{\affiliation{Department of Physics, The University of Texas at Dallas, Richardson, Texas 75080, USA}}
\newcommand{\israel}{\affiliation{Department of Particle Physics \& Astrophysics, Weizmann Institute of Science, Rehovot 7610001, Israel}}

\newcommand\orcid[1]{\href{https://orcid.org/#1}{$\!$\includegraphics[scale=0.006]{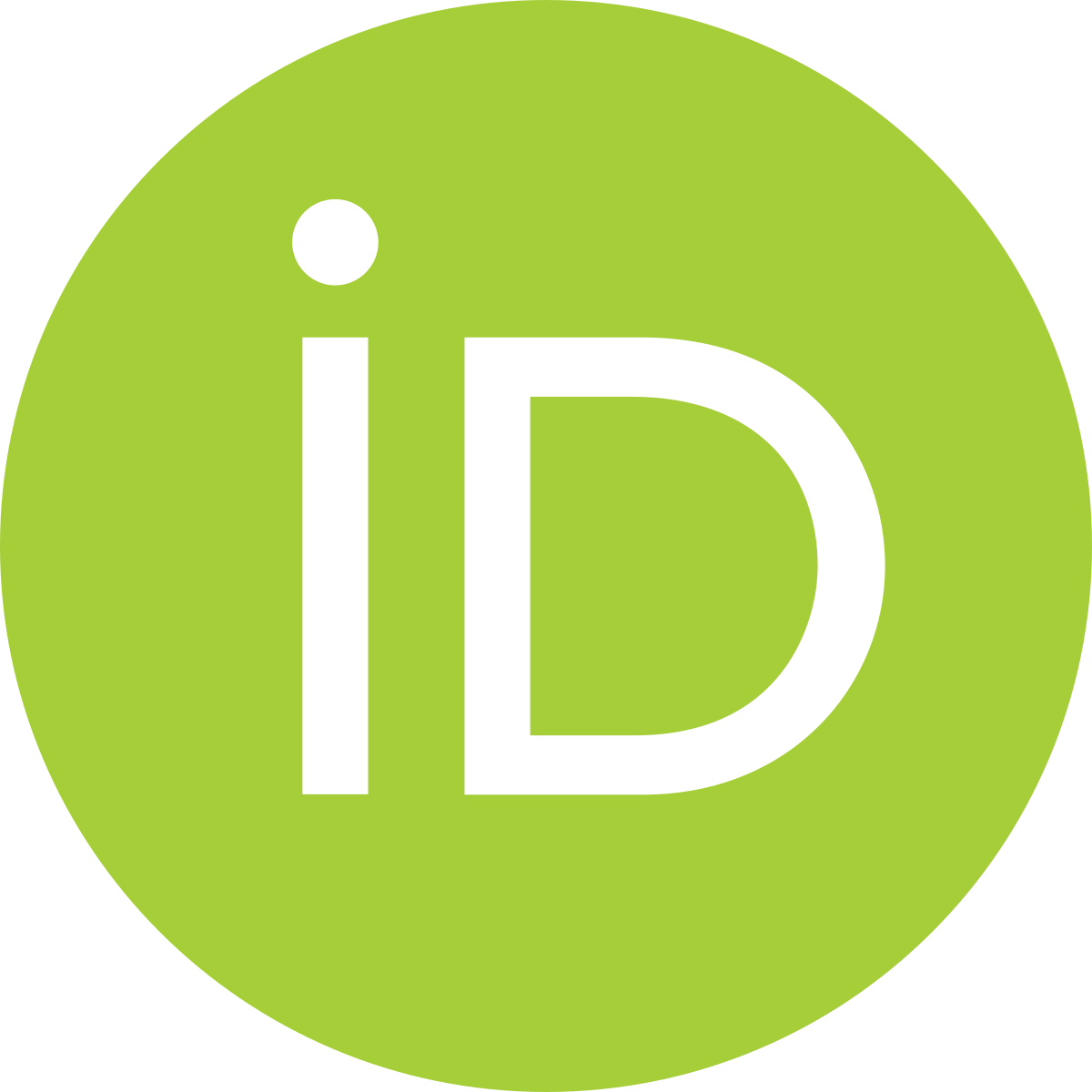} $\!\!$}}

\begin{document}

\title{Identifying multiple images of gravitational-wave sources lensed by elliptical lensing potentials}

\author{Saif Ali \orcid{0000-0002-6971-4971}}
\email{saif.ali@weizmann.ac.il}
\israel
\dallas

\author{Michael Kesden \orcid{0000-0002-5987-1471}}
\email{kesden@utdallas.edu}
\dallas

\author{Lindsay King \orcid{0000-0001-5732-3538}}
\email{lindsay.king@utdallas.edu}
\dallas

\date{\today}

\begin{abstract}
Real astrophysical lenses typically lack axisymmetry, necessitating the study of gravitational-wave (GW) lensing by elliptical mass distributions to accurately assess detectability and waveform interpretation. We investigate strong lensing using the singular isothermal ellipsoid (SIE) model, which produces two or four images depending on the source's position relative to lens caustics. Employing a quasi-geometrical optics framework, we determine that the geometrical-optics approximation holds reliably for lens masses above approximately $10^5 \, M_\odot$ at GW frequencies relevant for ground-based detectors $(\sim 10^2 \,\text{Hz})$, though wave-optics effects become significant for lower masses or sources near caustics. Our waveform mismatch analysis demonstrates that the use of three-image templates significantly improves our ability to distinguish source signals, reducing mismatches from $\mathcal{O}(10^{-1})$ to $\mathcal{O}(10^{-2})$, typically by factors between 1.5 and 5 compared to the standard two-image template model. At lens masses above $10^7 \, M_\odot$, diffraction effects become negligible for ground-based detectors, resulting in an additional mismatch reduction by a factor of approximately three. These findings highlight the critical need for multi-image templates in GW searches to enhance detection efficiency and accuracy.
\end{abstract}

\maketitle
\graphicspath{ {./Figures} }

\section{\label{sec:introduction}Introduction}
Massive objects, such as galaxies or clusters of galaxies, warp spacetime in their vicinity, leading to the distortion of ray bundles from distant sources. In strong lensing, multiple images, or single highly distorted images, are formed. Weak lensing must be studied statistically, using tiny distortions of the shapes of distant galaxies. This gravitational lensing effect, a remarkable consequence of Einstein's theory of general relativity \cite{1936Sci....84..506E}, has been established as a powerful tool in astrophysics. Weak and strong lensing of electromagnetic (EM) sources allows us to understand the distribution of dark matter on galaxy and galaxy cluster scales \cite{2002ApJ...572...25D} and measure cosmological parameters \cite{2019RPPh...82l6901O, 2016A&ARv..24...11T}. Numerous examples of multiple images of distant EM sources have been observed, due to strong lensing by an intervening inhomogeneous gravitational field \cite{2010CQGra..27w3001B, 2002A&A...388..373B, 2006MNRAS.372.1425H}. 

The first detection of gravitational waves (GWs) in 2015 and the subsequent observation of 90 GW events from merging compact objects by the LIGO and Virgo collaboration has opened a new window on the Universe \cite{abbott2021gwtc-3}. GW astronomy has provided powerful insights into several problems of modern astrophysics such as the speed of GWs \cite{abbott2021tests}, the population of black holes \cite{2021-pop}, and the equation of state of nuclear matter \cite{abbott2018gw170817}. 

GWs can also be deflected by intervening massive objects \cite{1971NCimB...6..225L, 1974IJTP....9..425O, 1999PThPS.133..137N, PhysRevLett.80.1138, Takahashi_2003,Oguri:2018muv, Li:2018prc}. Although lensing of GW sources has not yet been observed, the detection of lensing signatures in GW events will become more probable as the sensitivities of the current LIGO-Virgo-KAGRA (LVK) network improve and when future detectors such as the Cosmic Explorer (CE) \cite{evans2021horizon}, the Einstein Telescope (ET) \cite{Maggiore:2019uih}, the Deci-Hertz Interferometer Gravitational Wave Observatory (DECIGO) \cite{Kawamura:2020pcg}, and the Laser Interferometer Space Antenna (LISA) \cite{Barausse:2020rsu} begin their respective missions. The detection of GW lensing could lead to several new astrophysical studies, including increasing our understanding of the elusive intermediate-mass black hole population \cite{Lai:2018rto}, primordial black holes \cite{Diego:2019rzc, Oguri:2020ldf}, and constraining the source and lens distributions \cite{2022ApJ...929....9X}. Follow-up studies of lensed GW events with EM observations could enable us to locate the host galaxy to sub-arcsecond precision \cite{Yu:2020agu, Hannuksela:2020xor} and lead to precision cosmography \cite{Sereno:2011ty, Liao:2017ioi, Cao:2019kgn, Li:2019rns, wempe2022lensing}. EM and GW lensing complement each other because EM provides spatial resolution while GW lensing provides temporal resolution.

There is a crucial difference between the lensing of GW and EM waves, namely the validity of the geometrical-optics (GO) approximation \cite{Takahashi_2003}. Since the wavelengths of EM waves are much smaller than the Schwarzschild radius of the lens, the approximation is valid for EM observations. However, the frequency floor at which the LVK detectors operate is much lower ($\sim 10$ Hz) than radio-astronomy frequencies and will be even lower for future space-based detectors.  This will result in the breakdown of the GO approximation depending on the lens properties and configuration.

In the limit where the GO approximation is valid, GW lensing can be further classified as macrolensing or microlensing depending on whether the time delay between the images is longer or shorter than the time the GW events spend in the detector's sensitivity band (approximately the inspiral time from the frequency floor to the merger frequency) \cite{PhysRevD.107.103023}. Macrolensing splits the signal into multiple images which can be detected as repeated events in the detector, whereas in microlensing the multiple time-delayed images overlap and interfere resulting in a single modulated event in the detector \cite{1971NCimB...6..225L, 1974IJTP....9..425O, 1999PThPS.133..137N}. Wave optics is important for lenses with mass $M_L \lesssim 10^8 (f/{\rm mHz})^{-1} M_\odot$ (corresponding to $\sim 10^5\, M_\odot$ for LVK frequencies \cite{1999PThPS.133..137N, Takahashi_2003, Takahashi_2004}) and for more massive lenses when the source is located near a caustic.

Many GW lensing studies have focused on singular axisymmetric lens models such as the point mass (PM) or singular isothermal sphere (SIS) that produce at most two images \cite{Takahashi_2003, 2021MNRAS.503.3326C, Ezquiaga_2021, PhysRevD.107.103023, PhysRevD.111.084019}. For example, these lens models were used in \cite{2021MNRAS.503.3326C} to show that diffraction effects can be significant for GWs in the LVK sensitivity band lensed by masses as low as $M_L \lesssim 10^2 M_\odot$. Investigating these models was critical to the development of GW lensing theory, similar to the earlier development of EM lensing theory. However, to model realistic gravitational lenses that lack axisymmetry, one must employ elliptical lens models more consistent with the majority of galaxies seen in the Universe \cite{1983ApJ...271..551O, 1984ApJ...284....1T, 1986ApJ...310..568B}. In this work, we use the singular isothermal ellipsoid (SIE) \cite{1994A&A...284..285K, 2003A&A...397..825A, 2010PASJ...62.1017O} model, which is a natural generalization of the SIS model. The SIE model can produce at most four images, which is consistent with EM observations \cite{2003MNRAS.341...13B, 2012AJ....143..120O, 2023arXiv231004494D, 2004ApJ...610..679R}. Some previous studies have employed elliptical lens models in their analysis. For example, PM and SIE lens models were employed in \cite{2021MNRAS.508.4869M}, where the authors studied galaxy-scale SIE lensing of GWs where each strongly lensed GW image is further affected by a population of PM microlenses. In \cite{more2022improved}, the authors developed an improved statistic for the detection of GW sources lensed by an SIE model that accounted for the phase difference and magnification ratios between lensed GWs images in addition to image time delays as previously proposed in \cite{2018arXiv180707062H}.

The lensed GW strain is obtained by multiplying the unlensed GW strain by an \textit{amplification factor} $F$ given by the diffraction integral \cite{Takahashi_2003}. For axisymmetric lenses such as the PM and SIS, analytical expressions exist for the diffraction integral that provide $F$ for the full range of frequencies from the low-frequency wave-optics regime to the high-frequency GO limit \cite{Takahashi_2003}. Some studies have calculated $F$ using techniques such as Levin’s algorithm \cite{2020PhRvD.102l4076G, 2008mgm..conf..807M}, Fast Fourier Transform convolution \cite{2018arXiv181009058G}, and Picard-Lefschetz theory \cite{2019arXiv190904632F, 2023MNRAS.525.2107J}. However, accurately computing $F$ in the wave-optics regime for complex lens models such as the SIE still remains a challenging task. 

The GO approximation generally holds at source positions for which $f\Delta t_d \gg 1$, where $f$ is the GW frequency and $\Delta t_d$ is time delay between the images. The wave-optics regime occurs in the other limit $f\Delta t_d \lesssim 1$. As discussed previously, calculating $F$ in the wave-optics regime for the SIE lens model is computationally challenging. We therefore consider the quasi-GO regime, which holds in the limit $f \Delta t_d \gtrsim 1$ \cite{Takahashi:2004mc}. The SIE lens model can produce one, two or four images depending on the location of the source with respect to the caustics. It produces four images when the source is located inside a diamond-shaped astroid caustic \cite{1994A&A...284..285K} whose area in the source plane (and thus the lensing probability) scales as the square of the lens ellipticity. For the case of four images, we employ the framework of the quasi-GO approximation to quantify the breakdown of the GO approximation. This breakdown is caused by the diffraction integral for a given image no longer being dominated by extrema of the time-delay surface.

In this study, we utilize the quasi-GO limit to assess the lens masses and source positions for which the GO approximation is valid in the SIE model. This approach yields a minimum lens mass $M_{L (\min)}$ as a function of source position $\vec{y}$ below which the geometrical-optics approximation ceases to hold. Specifically, for any lens mass $M_L$ that satisfies $M_L > \max \limits_{j} M_{L (\min)}^{(j)}$, where $j=\{1, 2\}$ (or $\{1, 2, 3, 4\}$) in the two-image (four-image) regime of the SIE lens, the GO approximation remains valid. Additionally, within the four-image regime of the SIE lens, we distinguish between macroimages and microimages based on the inspiral time of a typical LVK GW event. Finally, we investigate the distinguishability of a four-imaged GW lensing event by employing different template configurations in the microlensing regime with valid GO approximation.  Consideration of this microlensing regime is essential for four-imaged lensing events with their shorter time delays and will become increasingly important for future GW detectors with lower frequency floors.

We begin in Sec.~\ref{sec:Wave optics effects on gravitaitonal waves} with a review of the gravitational lensing of GWs, including a discussion of the mathematical framework of the quasi-GO approximation. The validity of this approximation for the singular isothermal sphere lens model is examined in Sec.~\ref{sec:Validity of quasi-geometrical approximation}. In Sec.~\ref{sec:Singular Isothermal Ellipsoid (SIE)}, we present the SIE lens model and establish the relationships between lens and image parameters, following the treatment in \cite{PhysRevD.107.103023}. Sec.~\ref{sec:Criteria for geometrical-optics approximation} explores the lens mass limit beyond which the GO approximation is valid. Sec.~\ref{sec:Classification of macroimages and microimages} classifies GW events within the tangential caustics as either macrolensed or microlensed. Finally, in Sec.~\ref{sec:Waveform mismatch: Unlensed and lensed templates}, we perform a mismatch analysis using template banks to identify four-imaged lensed GW events. Unless otherwise stated, we employ relativistic units where G = c = 1.

\section{Wave optics effects on gravitational waves} \label{sec:Wave optics effects on gravitaitonal waves}

In the frequency domain, the lensed waveform $\tilde{h}^L(f)$ is the product of the unlensed waveform $\tilde{h}(f)$ and the amplification factor
\begin{equation}\label{eq:dimensionless amp factor}
F(w) = \frac{w}{2 \pi i} \int d^2 \vec{x} \, \exp[i w T(\vec{x}, \vec{y})]\,,
\end{equation}
a complex-valued, frequency-dependent function that is calculated from the diffraction integral \cite{Takahashi_2004}.  The amplification factor is a function of the dimensionless frequency
\begin{align} \label{E:w}
w \equiv \frac{D_S}{D_{LS} D_L} \xi_0^2 (2 \pi f) (1+z_L) = 8 \pi M_{Lz} f\,,
\end{align}
and the dimensionless source position $\vec{y} \equiv \boldsymbol{\eta}D_L/\xi_0 D_S$.  Here $D_L, D_S,$ and $D_{LS}$ are the angular-diameter distances from observer to lens, from observer to source, and from lens to source, $\xi_0$ is a model-dependent normalization length, $z_L$ is the lens redshift, and $\boldsymbol{\eta}$ is the physical position in the source plane with respect to the optic axis.  The amplification factor $F$ is independent of the choice of normalization length $\xi_0$ because the scaling of the dimensionless lens position $\vec{x} \equiv \boldsymbol{\xi}/\xi_0$ over which the diffraction integral is performed, where $\boldsymbol{\xi}$ is the physical position in the lens plane with respect to the optic axis.  Eq.~(\ref{E:w}) implicitly defines the model-dependent redshifted lens mass $M_{Lz}$.

In Eq.~(\ref{eq:dimensionless amp factor}), the Fermat potential
\begin{align} \label{eq:dimensionless time delay}
T(\vec{x}, \vec{y}) &\equiv \frac{D_L D_{LS}}{D_S \xi_0^2 (1+z_L)} t_d(\vec{x}, \vec{y})\,, \notag \\
&= \frac{1}{2} |\vec{x} - \vec{y}|^2 - \psi(\vec{x}) - \phi_m(\vec{y})\,,
\end{align}
determines the time delay $t_d(\vec{x}, \vec{y})$ for a GW traveling from source position $\vec{y}$ that passes through the lens plane at position $\vec{x}$.  Here $\psi(\vec{x})$ is the dimensionless gravitational lensing potential and $\phi_m(\vec{y})$ is chosen such that the time delay minimized with respect to lens position $\vec{x}$ is zero. A GW passing through the lens plane at position $\vec{x}$ is sourced from position $\vec{y}$ given by the lens equation
\begin{equation}\label{eq:lens eqn}
    \vec{y} = \vec{x} - \nabla_x \psi(\vec{x})\,.
\end{equation}

\subsection{\label{subsec: geometrical optics regime} Geometric-optics (GO) approximation}

In the high-frequency limit, only stationary points of the Fermat potential $T(\vec{x}, \vec{y})$ contribute to the diffraction integral of Eq.~(\ref{eq:dimensionless amp factor}). These stationary points are associated with image location(s) $\vec{x}^{(j)}$ in the lens plane satisfying the lens equation
\begin{equation}
\nabla_x T(\vec{x}^{(j)}, \vec{y}) = \vec{x}^{(j)} - \vec{y} - \boldsymbol{\alpha}(\vec{x}^{(j)}) = 0\,,
\end{equation}
where $\boldsymbol{\alpha}(\vec{x}^{(j)}) = \nabla_x \psi(\vec{x}^{(j)})$ is the deflection angle.  The Fermat potential $T(\vec{x}, \vec{y})$ can be Taylor expanded around image position $\vec{x}^{(j)}$ in $\vec{\tilde{{x}}} \equiv \vec{x} - \vec{x}^{(j)}$ \cite{2004A&A...423..787T}
\begin{align}\label{eq:time delay expansion go}
    T(\vec{x}, \vec{y}) &= T^{(j)} + \frac{1}{2} \sum_{a,b} \partial_a \partial_b T(\vec{x}^{(j)}, \vec{y}) \tilde{x}_a \tilde{x}_b + \mathcal{O}({\tilde{x}}^3),
\end{align}
where $T^{(j)} \equiv T(\vec{x}^{(j)}, \vec{y})$ and the terms linear in $\vec{\tilde{x}}$ vanish because $\vec{x}^{(j)}$ is the stationary point of $T(\vec{x}, \vec{y})$.  The indices $a, b \in \{1, 2\}$ denote the two components of the vector $\vec{x}^{(j)}$ on the lens plane. The amplification factor in the GO approximation can be obtained by inserting Eq.~(\ref{eq:time delay expansion go}) into Eq.~(\ref{eq:dimensionless amp factor}) \cite{1992grle.book.....S, 1999PThPS.133..137N, Matsunaga_2006}:
\begin{align}\label{eq:amplification factor go}
    F(w) = \sum_j |\mu^{(j)}|^{1/2} e^{\left(i w T^{(j)}- i \pi n^{(j)} \right)},
\end{align}
where the magnification of the $j$th image is given by $\mu^{(j)} = 1/{\rm det}(\partial\vec{y}/\partial\vec{x}^{(j)})$ and the Morse index $n^{(j)}$ has values of 0, 1/2, or 1 depending on whether $\vec{x}^{(j)}$ is a minimum, saddle point, or maximum of the Fermat potential.  

\subsection{Quasi-geometrical optics approximation} \label{subsec: Corrected geometrical-optics regime}

The GO approximation underlying the amplification factor of Eq.~(\ref{eq:amplification factor go}) breaks down when contributions to the diffraction integral of Eq.~(\ref{eq:dimensionless amp factor}) beyond the neighborhood of the stationary points can no longer be neglected. Corrections to this approximation can be obtained by expanding the amplification factor $F(w)$ in powers of $1/w$.  Expanding the Taylor expansion of the Fermat potential in Eq.~(\ref{eq:time delay expansion go}) to fourth order in $\vec{\tilde{{x}}}$ yields \cite{2004A&A...423..787T}
\begin{align} \label{eq:time delay expansion c-go}
    T(\vec{x}, \vec{y}) &= T^{(j)} + \frac{1}{2!} \sum_{a, b} T_{ab}^{(j)} \tilde{x}_a \tilde{x}_b + \frac{1}{3!} \sum_{a, b, c} T_{abc}^{(j)} \tilde{x}_a \tilde{x}_b \tilde{x}_c \notag \\ & + \frac{1}{4!} \sum_{a, b, c, d} T_{abcd}^{(j)} \tilde{x}_a \tilde{x}_b \tilde{x}_c \tilde{x}_e + \mathcal{O}(\tilde{x}^5)\,,
\end{align}
where
\begin{align}
T_{ab}^{(j)} &\equiv \partial_a \partial_b T(\vec{x}^{(j)}, \vec{y})\,, \\
T_{abc}^{(j)} &\equiv \partial_a \partial_b \partial_c T(\vec{x}^{(j)}, \vec{y})\,, \\
T_{abcd}^{(j)} &\equiv \partial_a \partial_b \partial_c \partial_d T(\vec{x}^{(j)}, \vec{y})\,.
\end{align}
Since $T_{ab}$ is a real, symmetric $2 \times 2$ matrix, one can construct the matrix $A_{ab}$ whose columns are its eigenvectors; this matrix satisfies
\begin{equation}
\sum_{a,b} T_{ab}A_{ac}A_{bd} = \lambda_{(c)} \delta_{cd}\,,
\end{equation}
where $\lambda_{(c)}$ are the eigenvalues of $A_{ab}$. Changing variables from $\tilde{x}_a$ to $z_a \equiv \sum_b A^{-1}_{ab} \tilde{x}_b$ in Eq.~(\ref{eq:time delay expansion c-go}) yields
\begin{align}\label{eq:time delay expansion c-go diag}
    T(\vec{x}, \vec{y}) &= T^{(j)} + \frac{1}{2!} \sum_a \lambda_{(a)} z_a^2 + \sum_{a,b,c}M_{abc} z_a z_b z_c \notag \\ & +  \sum_{a,b,c,d}N_{abcd} z_a z_b z_c z_d + \mathcal{O}(z^5),
\end{align}
where 
\begin{align}
    M_{abc} &\equiv \frac{1}{3!} \sum_{d,e,f} T_{def}^{(j)} A_{da} A_{eb} A_{fc}, \\ 
    N_{abcd} &\equiv \frac{1}{4!} \sum_{e,f,g,h} T_{efgh}^{(j)} A_{ea} A_{fb} A_{gc} A_{hd}.
\end{align}
Inserting Eq.~(\ref{eq:time delay expansion c-go diag}) into Eq.~(\ref{eq:dimensionless amp factor}) and then expanding Eq.~(\ref{eq:dimensionless amp factor}) to linear order in $1/w$ yields
\cite{2004A&A...423..787T}
\begin{align}\label{eq:amplification factor cgo}
    F(w) = \sum_j |\mu^{(j)}|^{1/2} \left(1 + i\frac{\Delta^{(j)}}{w}\right) e^{\left(i w T^{(j)} - i \pi n^{(j)} \right)},
\end{align}
where $\Delta^{(j)}$ is a real number given by \cite{2004A&A...423..787T}
\begin{align}\label{eq:deltaj non-axis}
    \Delta^{(j)} &= \frac{15}{2}\left(\frac{M_{111}^2}{\lambda_{(1)} |\lambda_{(1)}|^2} + 3\frac{M_{111} M_{122}}{|\lambda_{(1)}|^2 \lambda_{(2)} } + 3\frac{M_{112} M_{222}}{\lambda_{(1)} |\lambda_{(2)}|^2} \right. \notag \\ & \left. + \frac{M_{222}^2}{\lambda_{(2)} |\lambda_{(2)}|^2}\right) - 3 \left(\frac{N_{1111}}{|\lambda_{(1)}|^2} + 2\frac{N_{1122}}{\lambda_{(1)} \lambda_{(2)}} + \frac{N_{2222}}{|\lambda_{(2)}|^2}\right).
\end{align}

The GO approximation breaks down when this linear term becomes significant, i.e. for $w \lesssim \Delta^{(j)}$.

\subsection{Gravitational waveforms} \label{subsec:gravitational waveform}

We use the post-Newtonian (PN) approximation to model the inspiral phase of the gravitational waveform generated during binary black hole (BBH) mergers \cite{PhysRevD.49.2658}. Although PN waveforms break down near merger, they are a reasonable description of the early inspiral during which significant lensing-induced modulation occurs. The unlensed waveform $\tilde{h}(f)$ is given by
\begin{align}\label{eq:gw waveform}
    \tilde{h}(f)= 
    \begin{cases}
    (A/D)\mathcal{M}^{5/6} f^{-7/6} e^{i \Psi(f)},&  0 < f < f_{\rm cut}\\
    0,              & f_{\rm cut} < f,
    \end{cases}
\end{align}
where $A$ is the GW amplitude, $D$ is the luminosity distance to the source, $\mathcal{M} \equiv \eta^{3/5}(1+z_S)M$ is the redshifted chirp mass of a BBH system at redshift $z_S$ with total mass $M = m_1 + m_2$ and symmetric mass ratio $\eta = m_1 m_2/M^2$, and $\Psi(f)$ is the GW phase.

We approximate the cut-off frequency $f_{\rm cut} = (6^{3/2} \pi M_{z_S})^{-1}$ to be twice the orbital frequency at the innermost stable circular orbit of a Schwarzschild BH of mass $M_{z_S}$. The GW phase $\Psi(f)$ to 1.5PN order is given by \cite{PhysRevD.49.2658}
\begin{align}\label{eq:gw phase}
    \Psi (f) &= 2 \pi f t_c - \phi_c - \frac{\pi}{4} + \frac{3}{4} (8 \pi \mathcal{M} f)^{-5/3} \notag \\ & \left[1 + \frac{20}{9} \left(\frac{743}{336} + \frac{11 \eta}{4}\right) x - 16 \pi x^{3/2}\right],
\end{align}
where $t_c$ and $\phi_c$ are the coalescence time and phase respectively and $x \equiv (\pi M_{z_S} f)^{2/3}$ is the PN expansion parameter.  

\section{Validity of quasi-geometrical approximation} \label{sec:Validity of quasi-geometrical approximation}

Before advancing to elliptical lensing models that can produce four images, we will assess the validity of the quasi-GO approximation using the singular isothermal sphere (SIS) lens \cite{2004A&A...423..787T}. The lensing potential for the SIS lens is $\psi(x) = x$; inserting this potential into Eqs.~(\ref{eq:dimensionless amp factor}) and (\ref{eq:dimensionless time delay}) yields the exact amplification factor is \cite{Matsunaga_2006}
\begin{align}\label{eq: amp factor sis analytical}
    F(f) &= e^{iw[y^2 + 2\phi_m(y)]/2} \sum_{n=0}^\infty \frac{\Gamma(1 + n/2)}{n!} \nonumber \\
    & \quad \quad \times (2we^{3i\pi/2})^{n/2} {}_{1}F_1 \left(1 + \frac{n}{2}, 1; -\frac{i}{2}w y^2\right)\,, 
\end{align}
where $\phi_m(y) = y + 1/2$ and ${}_{1}F_1$ is the confluent hypergeometric function of the first kind. In the GO approximation, the number of images formed by the SIS lens is determined by the location of the source. The amplification factor in the GO approximation is given by \cite{Takahashi_2003}
\begin{align}\label{eq:amp factor sis geo}
    F(f) = 
    \begin{cases}
    |\mu_+|^{1/2} - i |\mu_-|^{1/2} e^{2 \pi i f \Delta t_d},&  y < 1\\
    |\mu_+|^{1/2},              & y > 1\,,
    \end{cases}
\end{align}
where
\begin{subequations} \label{eq:SISip}
\begin{align}
    \mu_\pm &= \pm 1 + \frac{1}{y}\,, \label{E:SISmag} \\
    \Delta t_d &= 8M_{Lz}y\,. \label{E:SIStd}
\end{align}
\end{subequations}
One image is always produced at $x_+ = y + 1$, and for $y < 1$, a second image is formed at $x_- = y - 1$.  Eq.~(\ref{eq:deltaj non-axis}) implies that 
\begin{align}\label{eq:deltaj sis}
    \Delta^{\pm} = \frac{1}{8 |x_\pm|^2 (|x_\pm| - 1)} = \frac{\pm 1}{8 y (y \pm 1)^2}\,,
\end{align}
for the two images.
Eq.~(\ref{eq:amplification factor cgo}) indicates that for image $j$, the quasi-GO approximation is valid when
\begin{align}\label{eq:geo valid}
    \frac{\Delta^{(j)}}{w} \ll 1 \implies f \gg \frac{\Delta^{(j)}}{8 \pi M_{Lz}}\,.
\end{align}
Since this condition must be satisfied for \emph{all} images for the amplification factor of Eq.~(\ref{eq:amplification factor go}) to be valid, we require $w > \max \limits_{j} \{\Delta^{(j)} \}$.

\begin{figure} 
    \centering
    \includegraphics[scale = 0.37]{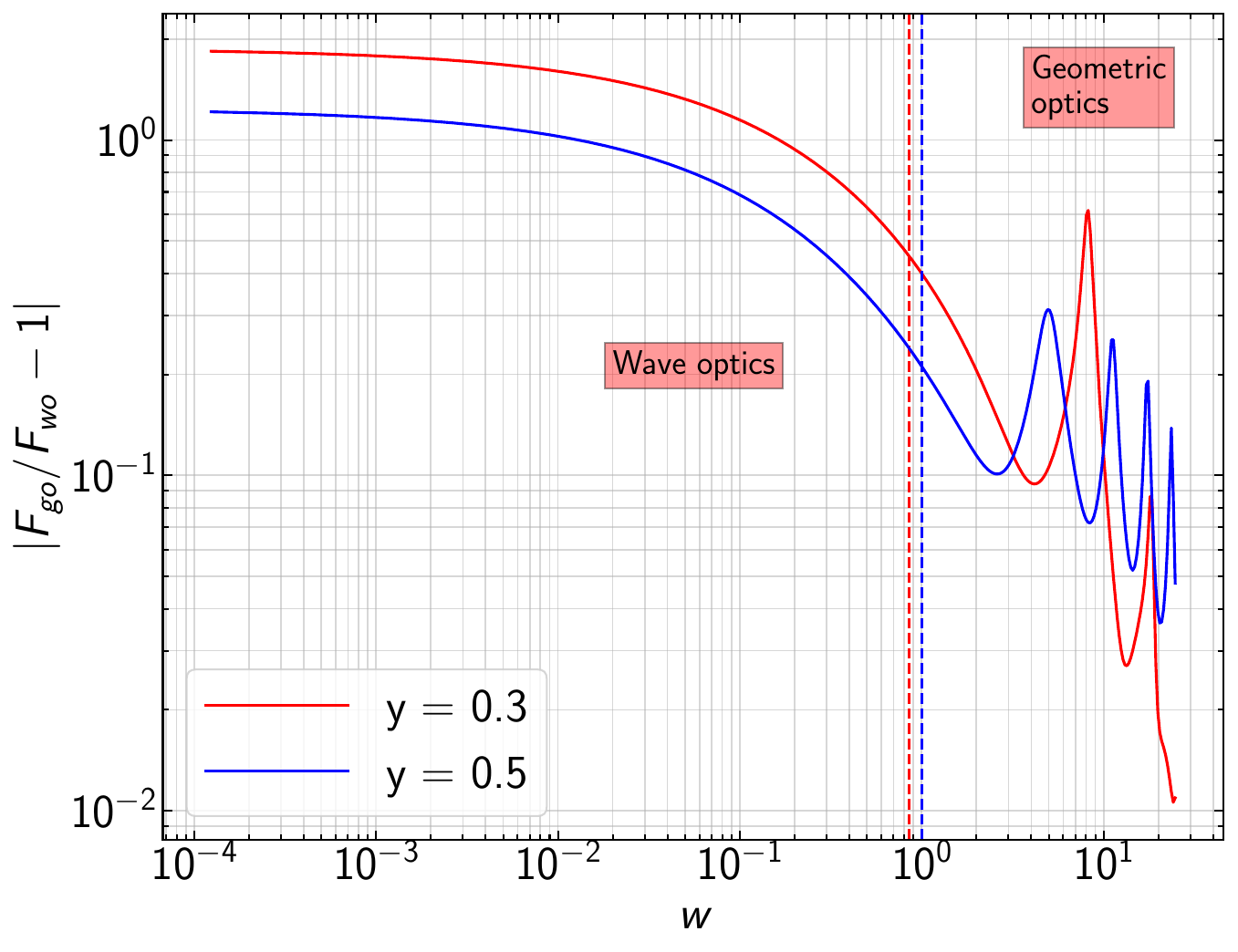}
    \caption{
    Fractional error in the amplification factor $F_{\rm go}$ in the GO approximation with respect to the true wave-optics result $F_{\rm wo}$ as a function of the dimensionless frequency $w = 8 \pi M_{Lz} f$. The solid red (blue) curves show this fractional error for dimensionless source positions $y = 0.3$ ($0.5$), while the corresponding dashed vertical lines $w = \max \limits_{j} \{\Delta^{(j)} \}$ indicate the frequencies above which the GO approximation is valid.
    }
    \label{fig:deltaj valid}
\end{figure}

In Fig.~\ref{fig:deltaj valid}, we plot the fractional error $|F_g/F_w - 1|$ in the GO approximation as a function of $w$. We observe that this fractional error is indeed less than unity for $w > \max \limits_{j} \{\Delta^{(j)} \}$ implying that the GO approximation is reasonably valid in this regime.

\begin{figure*}[t!] 
    \centering
    \includegraphics[width=\textwidth, height=11cm]{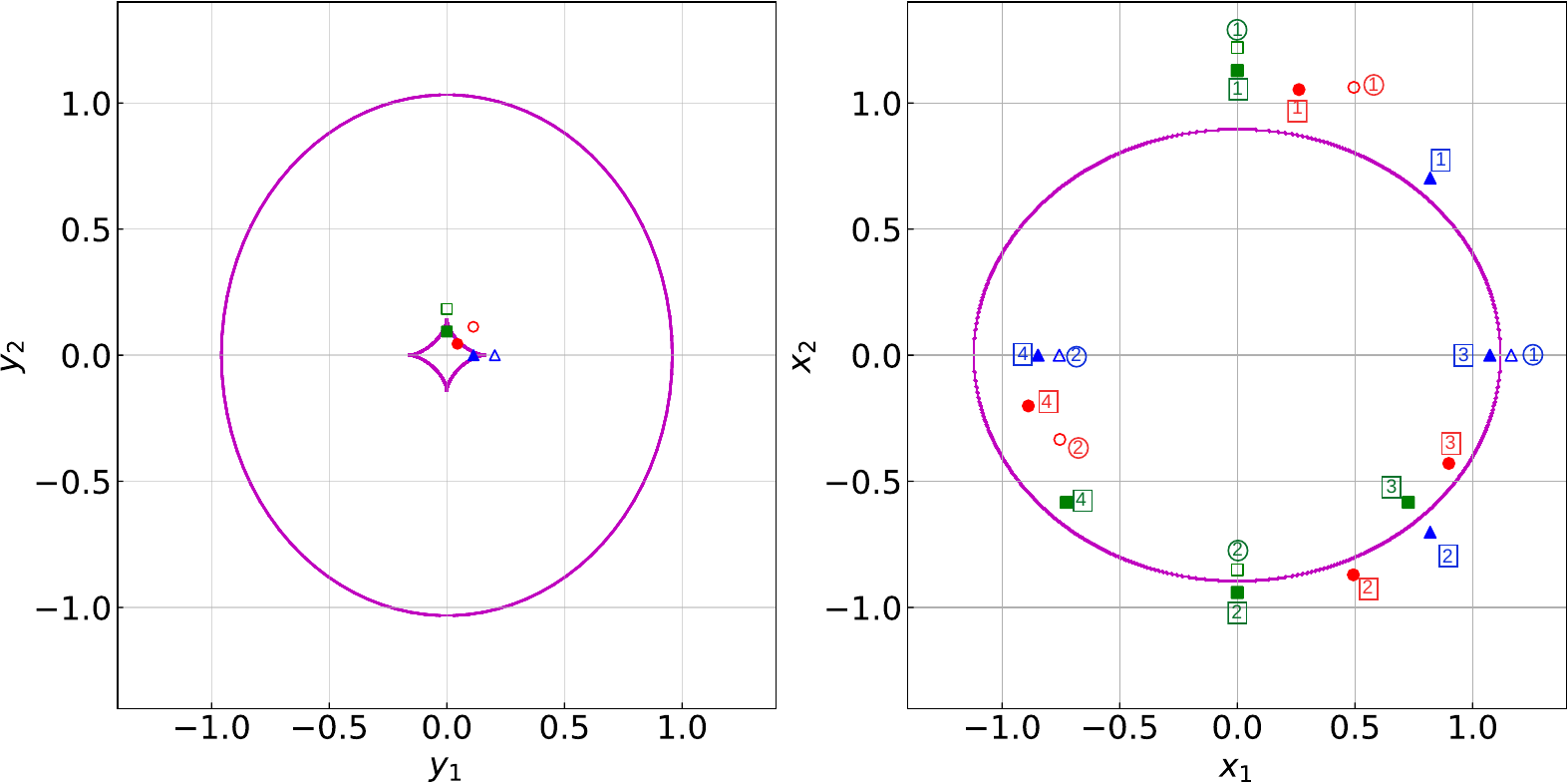}
    \caption{The left (right) panel shows the source (image) plane and the caustics (critical curves) for a SIE lens with ellipticity $e = 0.2$. The blue triangles, red circles, and green squares correspond to source angles $\phi = 0$ (major cusp), $45^\circ$ (fold), and $90^\circ$ (minor cusp), while the empty (solid) symbols correspond to source positions $y$ outside (inside) the inner caustic leading to two (four) images in the GO approximation. The number enclosed by the square (circle) denotes the sequence of arrival times for the images in the two-image (four-image) regime.
    }
    \label{fig:caustics sigma=6}
\end{figure*}

\section{\label{sec:Singular Isothermal Ellipsoid (SIE)}Singular Isothermal Ellipsoid (SIE)}

As real gravitational lenses are not axisymmetric, we will consider an elliptical lens models that is consistent with the majority of galaxies seen in the Universe \cite{1983ApJ...271..551O, 1984ApJ...284....1T, 1986ApJ...310..568B}. Elliptical lens models have mass distributions such that contours of constant surface mass density are ellipses. In this work, we will use the singular isothermal ellipsoid (SIE) model \cite{1994A&A...284..285K, 2003A&A...397..825A, 2010PASJ...62.1017O} which is a natural generalization of the SIS model. It breaks the axisymmetry about the optic axis present in lens models such as point mass (PM) and SIS lens and provides a good fit to the early-type galaxies that dominate strong lensing \cite{1984ApJ...284....1T, 2007MNRAS.379.1195M}.

The SIE lensing potential is \cite{2010PASJ...62.1017O}
\begin{align}\label{eq:lens potential}
    \psi(\vec{x}) &= \sqrt{\frac{q}{1-q^2}} \left[x_1 \tan^{-1}\left(\frac{\sqrt{1-q^2}x_1}{\sqrt{q^2x^2_1+x^2_2}} \right) \right. \notag \\ & \quad \left. + x_2 \tanh^{-1} \left(\frac{\sqrt{1-q^2}x_2}{\sqrt{q^2x^2_1+x^2_2}}\right) \right]
\end{align}
where the axis ratio for ellipticity $e$ is defined as $q \equiv 1-e$.
In the limit $e \to 0$, this lensing potential reduces to the axisymmetric SIS result $\psi(\vec{x}) = x$. For axisymmetric lens models such as the SIS, the lens parameters consist of the source position $y$ and the redshifted lens mass $M_{Lz}$ inside the Einstein radius $\xi_0$. The SIE introduces two new parameters: the ellipticity $e$ and the angle $\phi$ between the source position $\vec{y}$ and the major axis of the ellipse.

\begin{figure*}[t!]
    \centering
    \includegraphics[height=20cm, width=\textwidth]{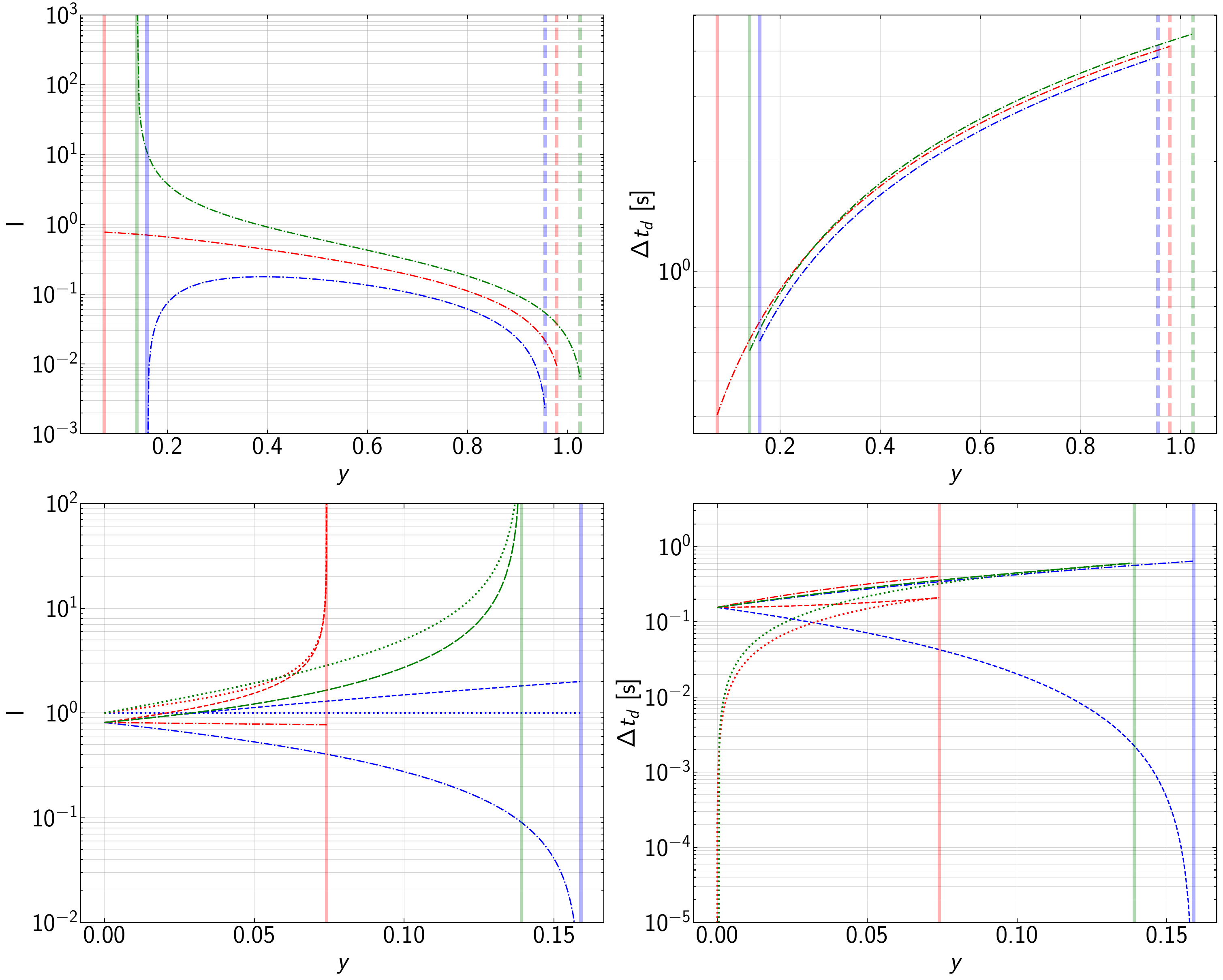}
    \caption{Flux ratios $I^{(j)}$ (left panels) and time delays $\Delta t_d^{(j)}$ (right panels) in the two-image (top panels) and four-image (bottom panels) regimes as a function source position $y$ for a SIE lens with redshifted lens mass $M_{Lz}= 1.05 \times 10^5 M_\odot$ and ellipticity $e = 0.2$. The blue, red, and green curves correspond to source angles $\phi = 0^\circ$, $45^\circ$, and $90^\circ$, respectively. The dash-dotted curves show the single flux ratio and time delay in the two-image regime (top panels), while the dotted, dashed, and dash-dotted curves represent the flux ratios and time delays for the second, third, and fourth images, respectively, in the four-image regime (bottom panels). Note that for the major-cusp case ($\phi = 0^\circ$), symmetry implies that the first and second images have the same magnification and arrival time, so $I^{(2)} = 1$ and $\Delta t_d^{(2)} = 0$.  For the minor-cusp case ($\phi = 90^\circ$), symmetry implies that the third and fourth images have the same magnification and time delay, so $I^{(3)} = I^{(4)}$ and $\Delta t_d^{(3)} = \Delta t_d^{(4)}$. The thick solid and dashed horizontal lines indicate the locations $y_t(\phi)$ and $y_r(\phi)$ of the inner (tangential) and outer (radial) caustics.}
    \label{fig:image param vs y images}
\end{figure*}

Fig.~\ref{fig:caustics sigma=6} shows the caustics (left panel) and critical curves (right panel) for a SIE lens. The outer caustic in the left panel is smooth and known as a `pseudo-caustic' or radial caustic. The inner caustic is an astroid consisting of four smooth concave segments called folds that intersect at cusps; it is known as a `true caustic' or tangential caustic \cite{2000ApJ...537..697K}. The source position $\vec{y}$ with respect to these caustics determines the number of images produced: sources outside the outer (radial) caustic [$y > y_r(\phi)$], between the two caustics [$y_t(\phi) < y < y_r(\phi)$], and inside the inner (tangential) caustic [$y < y_t(\phi)$] produce one, two, and four images respectively. We choose three different source positions: alignment with the major cusp $(\phi = 0^\circ)$, fold $(\phi = 45^\circ)$, and minor cusp $(\phi = 90^\circ)$.  Without loss of generality, we restrict our analysis to the first quadrant of the source plane. We use \textsc{glafic} \cite{2010PASJ...62.1017O} to calculate the image positions, magnifications, and Morse indices as functions of the source position $\vec{y}$ and ellipticity $e$.

\subsection{\label{subsec:Lens and image parameters} Lens and image parameters}

For axisymmetric, two-image lens models, the two image parameters, the flux ratio $I \equiv |\mu_-|/|\mu_+|$ and time delay $\Delta t_d$, are model-dependent functions of the two lens parameters $y$ and $M_{Lz}$ \cite{PhysRevD.107.103023}.  For lens models for which these functions are monotonic, they can be inverted to allow observational constraints on the image parameters to be propagated into constraints on the lens parameters. In the case of the SIE lens, we have four lens parameters, $\{ y, \phi, e, M_{Lz} \}$, and two (six) image parameters $\{ I^{(j)}, \Delta t_d^{(j)} \}$ for $j = 2\,(4)$ in the two- (four-) image regime.  The flux ratios are normalized by the magnification of the first image and ordered with respect to the first image which is defined to have zero time delay. These image parameters for GW lensing differ from those of most EM lensing, where static sources prevent time delays from being measured but superior spatial resolution allows image positions to be determined.

In Fig.~\ref{fig:image param vs y images}, we plot the image parameters $I^{(j)}$ and $\Delta t_d^{(j)}$ as functions of $y$ in the two-image and four-image regimes for a SIE lens with fiducial redshifted lens mass $M_{Lz} = 1.05 \times 10^5 M_\odot$ and eccentricity $e = 0.2$. The top left panel shows $I^{(2)}$ as a function of $y$ for three different source angles ($\phi = 0^\circ,\, 45^\circ, \, 90^\circ$) in the two-image regime $y_t(\phi) < y < y_r(\phi)$. For all three source angles, the magnification of the second image $\mu^{(2)}$ vanishes as $y \rightarrow y_r$ implying $I^{(2)} \rightarrow 0$ in this limit. In the other limit $y \rightarrow y_t$, we observe differing behavior for the three source angles: the first (second) image approaches the critical curve for $\phi = 0^\circ~(90^\circ)$ leading to $I^{(2)} \rightarrow 0~(\infty)$; for the fold case $\phi = 45^\circ$, neither image approaches the critical curve, so both magnifications remain finite and $I^{(2)} \rightarrow 0.8$.  The top right panel of Fig.~\ref{fig:image param vs y images} shows the time delay $\Delta t_d^{(2)}$ between the two images as a function of $y$.  It only depends weakly on the source angle $\phi$ and is well approximated by the SIS result
\begin{equation}
\Delta t_d^{\rm SIS} = 8M_{Lz}y = 4.2\,{\rm s} \left( \frac{M_{Lz}}{1.05 \times 10^5 M_\odot} \right)y~.
\end{equation}

In the bottom left and right panels of Fig.~\ref{fig:image param vs y images}, we plot the flux ratios $I^{(j)}$ and time delays $\Delta t_d^{(j)}$ as functions of source position $y$ for source angles $\phi$ in the four-image regime $y < y_t(\phi)$.  We begin by discussing the limiting behavior as $y \to y_t$. For $\phi = 0^\circ$, images 1 and 2 have identical magnifications by symmetry as shown in Fig.~\ref{fig:caustics sigma=6}.  Images 1, 2, and 3 all approach the critical curve in this limit while image 4 does not, implying that $I^{(2)} \to 1$, $I^{(3)} \to 2$, and $I^{(4)} \to 0$.  Symmetry implies that $\Delta t_d^{(2)} = 0$ and $\Delta t_d^{(3)} \to 0$ in this limit.  For $\phi = 90^\circ$, images 3 and 4 have identical magnifications by symmetry and images 2, 3, and 4 approach the critical curve, yielding $I^{(2)} \to \infty$ and $I^{(3)} = I^{(4)} \to I^{(2)}/2$ and $\Delta t_d^{(3)} = \Delta t_d^{(4)} \to \Delta t_d^{(2)} \neq 0$.  Finally, for $\phi = 45^\circ$, symmetry is broken but images 2 and 3 still approach the critical curve as $y \to y_t$ implying $I^{(2)} \to I^{(3)} \to \infty$ and $I^{(4)} \to {\rm const}$, as well as $\Delta t_d^{(2)} \to \Delta t_d^{(3)} \neq \Delta t_d^{(4)}$. In the opposite limit $y \to 0$, the four images form an Einstein cross with $|\mu^{(1)}| = |\mu^{(2)}|$ and $|\mu^{(3)}| = |\mu^{(4)}|$.  This implies $I^{(2)} \to 1$ and $I^{(3)} \to I^{(4)}$, as well as $\Delta t_d^{(2)} \to 0$ and $\Delta t_d^{(3)} \to \Delta t_d^{(4)}$.

\section{\label{sec:Criteria for geometrical-optics approximation} Criteria for geometrical-optics approximation}

\begin{figure} 
    \centering
    \includegraphics[scale=0.38]{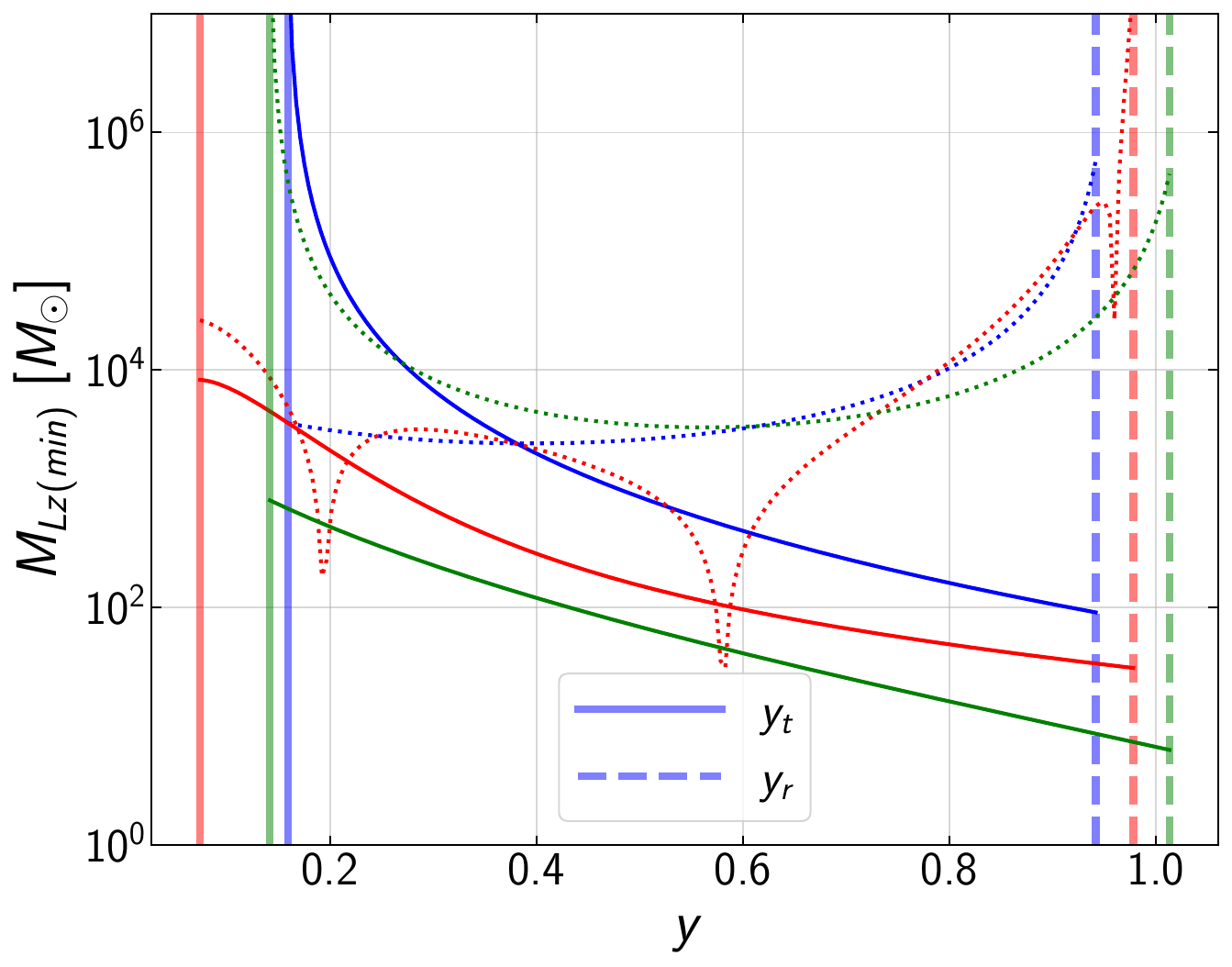}
    \caption{Minimum redshifted lens mass $M_{Lz ({\rm min})}^{(j)}$ in the two-image regime for ellipticity $e=0.2$ and minimum GW frequency $f_{\rm min} = 20$\,Hz as a function of source distance $y$.  The solid and dotted curves correspond to images 1 and 2 respectively.  As in Fig.~\ref{fig:image param vs y images}, the blue, red, and green curves correspond to source angles $\phi = 0^\circ$, $45^\circ$, and $90^\circ$, while the thick solid and dashed horizontal lines indicate the locations $y_t(\phi)$ and $y_r(\phi)$ of the inner (tangential) and outer (radial) caustics.
    }
    \label{fig:deltaj ML two_imgs}
\end{figure}

In Sec.~\ref{sec:Validity of quasi-geometrical approximation}, we presented the quasi-GO approximation \cite{2004A&A...423..787T} which provides the correction to the GO regime to order $\mathcal{O}(w^{-1})$. This correction is proportional to the term $\Delta^{(j)}$ given by Eq.~(\ref{eq:deltaj non-axis}). It follows that for the GO approximation to be valid, all the images must satisfy Eq.~(\ref{eq:geo valid}).  We use this prescription to explore the validity of the GO approximation in the two- and four-image regions of the lens parameters space.

Since the GW frequency $f$ of compact binaries increases as they inspiral towards merger, Eq.~(\ref{eq:geo valid}) implies that the GO approximation will be valid throughout a GW event if the redshifted lens mass is above the minimum value
\begin{equation}\label{eq:Mlz min}
M_{Lz ({\rm min})}^{(j)} = \frac{|\Delta^{(j)}|}{8\pi f_{\rm min}} \,,
\end{equation}
for all images $j$, where $f_{\rm min} = 20$\,Hz is the floor of the sensitivity band of the ground-based GW detector network.

In Fig.~\ref{fig:deltaj ML two_imgs}, we plot $M_{Lz ({\rm min})}^{(j)}$ as a function of source position $y$ in the two-image regime of the SIE lens. $M_{Lz ({\rm min})}^{(2)} \rightarrow \infty$ as $y \rightarrow y_r$ for all image configurations, however the GO approximation of Eq.~(\ref{eq:amplification factor cgo}) remains valid since the product $|\mu^{(2)}|^{1/2}\Delta^{(2)} \to 0$ in this limit. In the opposite limit $y \rightarrow y_t$, we see that $M_{Lz (min)}^{(1)}\,(M_{Lz (min)}^{(2)})\rightarrow \infty$ for $\phi = 0^\circ$ ($\phi = 90^\circ$), consistent with image 1 (2) approaching the critical curve.  For $\phi = 45^\circ$, $M_{Lz ({\rm min})}^{(j)}$ remains finite for both images as neither approaches the critical curve.  We conclude that for any redshifted lens mass $M_{Lz}$, the quasi-GO approximation remains valid for any source positions $\vec{y}$ that satisfy
\begin{equation} \label{E:QGOcond}
M_{Lz} > \max\limits_{j} M_{Lz ({\rm min})}^{(j)}(\vec{y})\,.
\end{equation}
This range includes most of the two-image regime $y_t(\phi) < y < y_r(\phi)$ for $M_{Lz} \gtrsim 10^4 M_\odot$ for ground-based GW detectors with $f_{\rm min} = 20$\,Hz.

\begin{figure*}[t!]
    \centering
    \includegraphics[scale = 0.34]{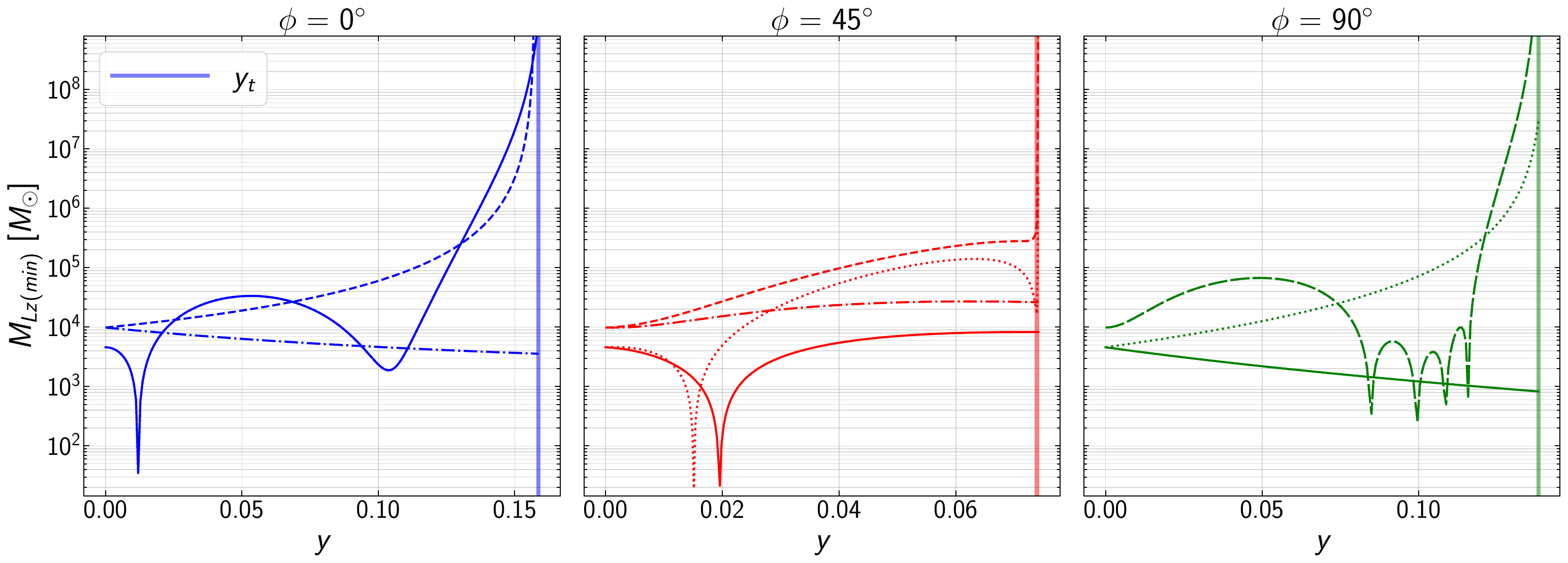}
    \caption{Minimum redshifted lens mass $M_{Lz ({\rm min})}$ in the four-image regime for ellipticity $e=0.2$ and minimum GW frequency $f_{\rm min} = 20$\,Hz as a function of source distance $y$. The three panels represent cusp and fold configurations with $\phi = 0^\circ$ (left), $45^\circ$ (middle), $90^\circ$ (right). The solid, dotted, dashed, and dot-dashed curves correspond to the first, second, third, and fourth images respectively.  As symmetry implies that the first and second (third and fourth) images are identical for $\phi = 0^\circ$ ($\phi = 90^\circ$), we do not show the dotted (dot-dashed) curves in the left (right) panel.  The vertical line represents the tangential caustic $y_t(\phi)$.}
    \label{fig:deltaj ML four_imgs}
\end{figure*}

In Fig.~\ref{fig:deltaj ML four_imgs}, we show $M_{Lz ({\rm min})}^{(j)}$ as a function of $y$ in the four-image regime. This minimum redshifted lens mass is continuous for images that are unaffected as the source crosses the inner caustic at $y_t(\phi)$.  Thus,
\begin{align}
M_{Lz ({\rm min})}^{(1)}(y \to y_t^-) &= M_{Lz ({\rm min})}^{(1)}(y \to y_t^+) \quad \quad \phi \neq 0^\circ \nonumber \\
M_{Lz ({\rm min})}^{(4)}(y \to y_t^-) &= M_{Lz ({\rm min})}^{(2)}(y \to y_t^+)  \quad \quad \phi \neq 90^\circ \nonumber
\end{align}
since the second and third images in the four-image regime emerge with shorter time delays than the previous second image which becomes the fourth image.  The images located at the critical curve (the second and third images as well as the first (fourth) images for $\phi = 0^\circ$ ($\phi = 90^\circ$)) all have $M_{Lz ({\rm min})}^{(j)} \to \infty$ as $y \to y_t$.  In the limit $y \to 0$, the images form an Einstein cross for which $M_{Lz ({\rm min})}^{(1)} = M_{Lz ({\rm min})}^{(2)}$ and $M_{Lz ({\rm min})}^{(3)} = M_{Lz ({\rm min})}^{(4)}$.  As gradients in the lensing potential are steeper in the four-image regime, larger redshifted lens masses $M_{Lz} \gtrsim 10^5 M_\odot$ are generally required for the GO approximation to remain valid.

\begin{figure*}[t!]
    \includegraphics[scale = 0.34]{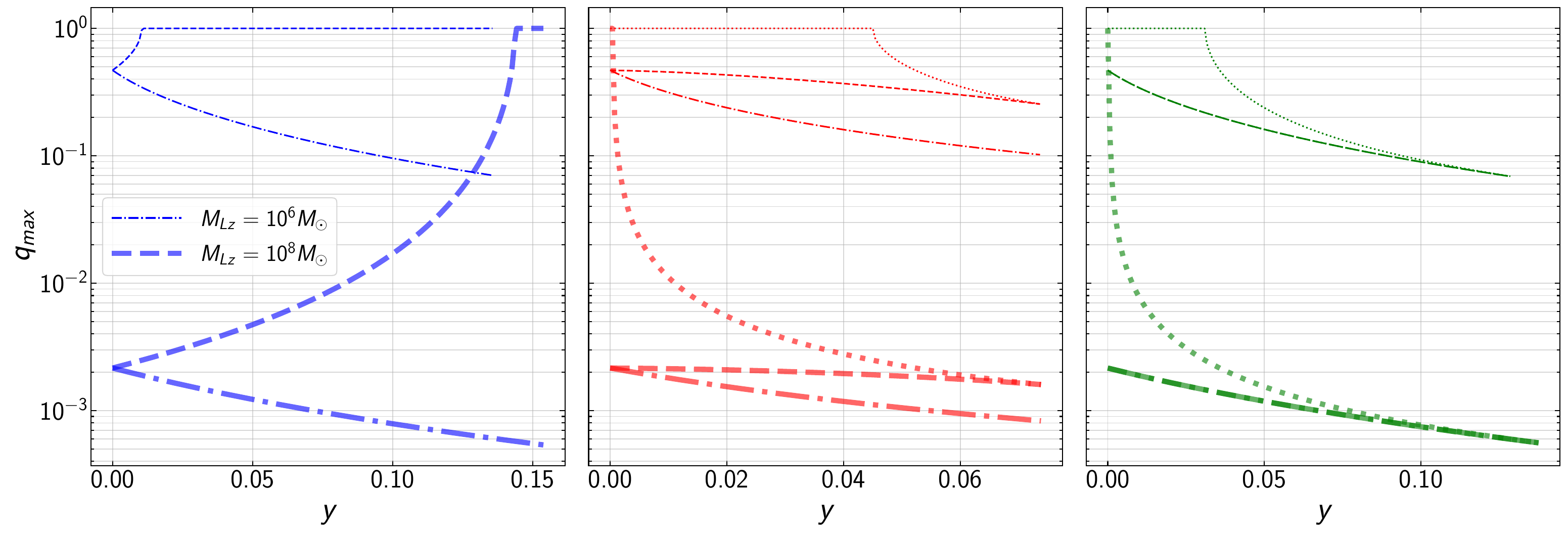}
    \caption{Maximum mass-ratio $q_{\rm max}$ is shown as a function of the source distance $y$ for BBH with total mass $M_{zS} = 50 M_\odot$. The three panels shows the cusp and fold configurations with $\phi = 0^\circ$ (left), $\phi= 45^\circ$ (middle), and $\phi=90^\circ$ (right). The thin (thick) curve corresponds to the $M_{Lz} = 10^6 M_\odot$ $(10^8 M_\odot)$. The dotted, dashed, and dot-dashed lines correspond to the second, third, and fourth images respectively.
    }
    \label{fig:q max}
\end{figure*}

\section{\label{sec:Classification of macroimages and microimages} Classification of macroimages and microimages}

In the previous section, we used the quasi-GO approximation to determine the minimum redshifted lens mass $M_{Lz ({\rm min})}$ above which the GO approximation is valid for all images. In this section, we will investigate, for lens masses above this value, under what conditions multiple images appear in the sensitivity band of the GW detector simultaneously (microlensing) versus sequentially (macrolensing). We borrow this terminology from EM lensing which is described as macrolensing or microlensing depending on whether the multiple images can or cannot be spatially resolved.
To lowest PN order, a GW signal spends a time \cite{PhysRevD.49.2658}
\begin{align} \label{eq:t_band}
\tau &= t(f_{\rm min}) - t(f_{\rm cut}) \nonumber \\
     &=  \frac{5 M_{z_S}}{\eta} \left[ \left(8 \pi M_{z_S} f_{\rm min} \right)^{-8/3} - \frac{81}{16} \right] \nonumber \\
     &\approx 1.29\,{\rm s}\,(4\eta)^{-1} \left( \frac{M_{z_S}}{50\,M_\odot} \right)^{-5/3} \left( \frac{f_{\rm min}}{20\,{\rm Hz}} \right)^{-8/3} 
\end{align}
in band for our default choices $f_{\rm min} = 20$~Hz and $f_{\rm cut} = (6^{3/2} \pi M_{z_S})^{-1}$. Equating $\tau$ to the time delay $\Delta t_d^{(j)}$ of the $j$th image determines the maximum mass ratio $q_{{\rm max}}^{(j)}$ for which the first and $j$th image are in band simultaneously.

\begin{table*}[t!]
    \centering
    \begin{tabular}{ |c||c|c|c|c|c|c|c||c|c|c|c|c|c|}
    \hline
     Class & $\Delta t_d^{(12)}$ & $\Delta t_d^{(13)}$ & $\Delta t_d^{(14)}$ & $\Delta t_d^{(23)}$ & $\Delta t_d^{(24)}$ & $\Delta t_d^{(34)}$ & Transitions &  &  &  &  &  &  \\
     \hline 
     Lens mass &  &  &  &  &  &  &                        & 0.1 & 0.5 & 0.81 & 2.3 & 3.5 & 10 \\ \hline \hline
     WO &  &  &  &  &  &  &                               & 59.38 & 22.23 & 13.81 & 7.87 & 5.69 & 2.32 \\ \hline 
     1 & + & + & + & + & + & + & 2                        & 40.62 & 13.32 & 0.13 & 0 & 0 & 0 \\ \hline
     2 & + & + & -- & + & + & + & 1, 3a                   & 0 & 30.7 & 1.81 & 0 & 0 & 0 \\ \hline
     3a & + & -- & -- & + & + & + & 2, 4a, 4b             & 0 & 5.03 & 10.61 & 0 & 0 & 0 \\ \hline
     3b & + & + & -- & + & -- & + & 2, 4b, 4c             & 0 & 4.63 & 10.24 & 0 & 0 & 0 \\ \hline
     4a & -- & -- & -- & + & + & + & 3a, 5a               & 0 & 12.7 & 29 & 9.12 & 6.14 & 1.59 \\ \hline
     4b & + & -- & -- & + & -- & + & 3a, 3b, 5a, 5b, 5c   & 0 & 0 & 1.57 & 0 & 0 & 0 \\ \hline
     4c & + & + & -- & + & -- & -- & 3b, 5c               & 0 & 11.39 & 28.1 & 9.66 & 6.8 & 1.51 \\ \hline
     5a & -- & -- & -- & + & -- & + & 4a, 4b, 6a, 6b      & 0 & 0 & 0.77 & 5.62 & 3.63 & 0.99 \\ \hline
     5b & + & -- & -- & -- & -- & + & 4b, 6a, 6c          & 0 & 0 & 0 & 2.82 & 1.51 & 0.22 \\ \hline
     5c & + & -- & -- & + & -- & -- & 4b, 4c, 6b, 6c      & 0 & 0 & 0.85 & 5.63 & 3.71 & 1.04 \\ \hline
     6a & -- & -- & -- & -- & -- & + & 5a, 5b, 7          & 0 & 0 & 0 & 7.13 & 8.29 & 5.95 \\ \hline
     6b & -- & -- & -- & + & -- & -- & 5a, 5c, 7          & 0 & 0 & 3.11 & 40.15 & 41.99 & 32.04 \\ \hline
     6c & + & -- & -- & -- & -- & -- & 5b, 5c, 7          & 0 & 0 & 0 & 7.38 & 8.42 & 5.9 \\ \hline
     7 & -- & -- & -- & -- & -- & -- & 6a, 6b, 6c         & 0 & 0 & 0 & 4.62 & 13.82 & 48.42 \\ \hline
    \end{tabular}
    \caption{Image classification for four-image lenses like the singular isothermal ellipsoid (SIE) model with source positions inside its tangential caustic.  The first column lists the wave-optics (WO) regime and the 14 classes in the GO regime.  Columns 2 through 7 show the six different microlensing conditions $\Delta t_d^{(jk)} \equiv \Delta t_d^{(k)} - \Delta t_d^{(j)}$; each is satisfied (+) for a given class if $\Delta t_d^{(jk)} < \tau$ for event duration $\tau$.  Column 8 lists the possible classes to which an event of a given class with a fixed dimensionless source position $\vec{y}$ can transition as the lens mass $M_{Lz}$ changes.  Columns 9 through 14 give the percentage of events in the WO regime and 14 classes for SIE lenses with $e = 0.2$ and masses as indicated in units of $10^6 M_\odot.$}
    \label{tab:classes}
\end{table*}

\begin{figure*}[t!]
        \includegraphics[width=0.81\textwidth, keepaspectratio]{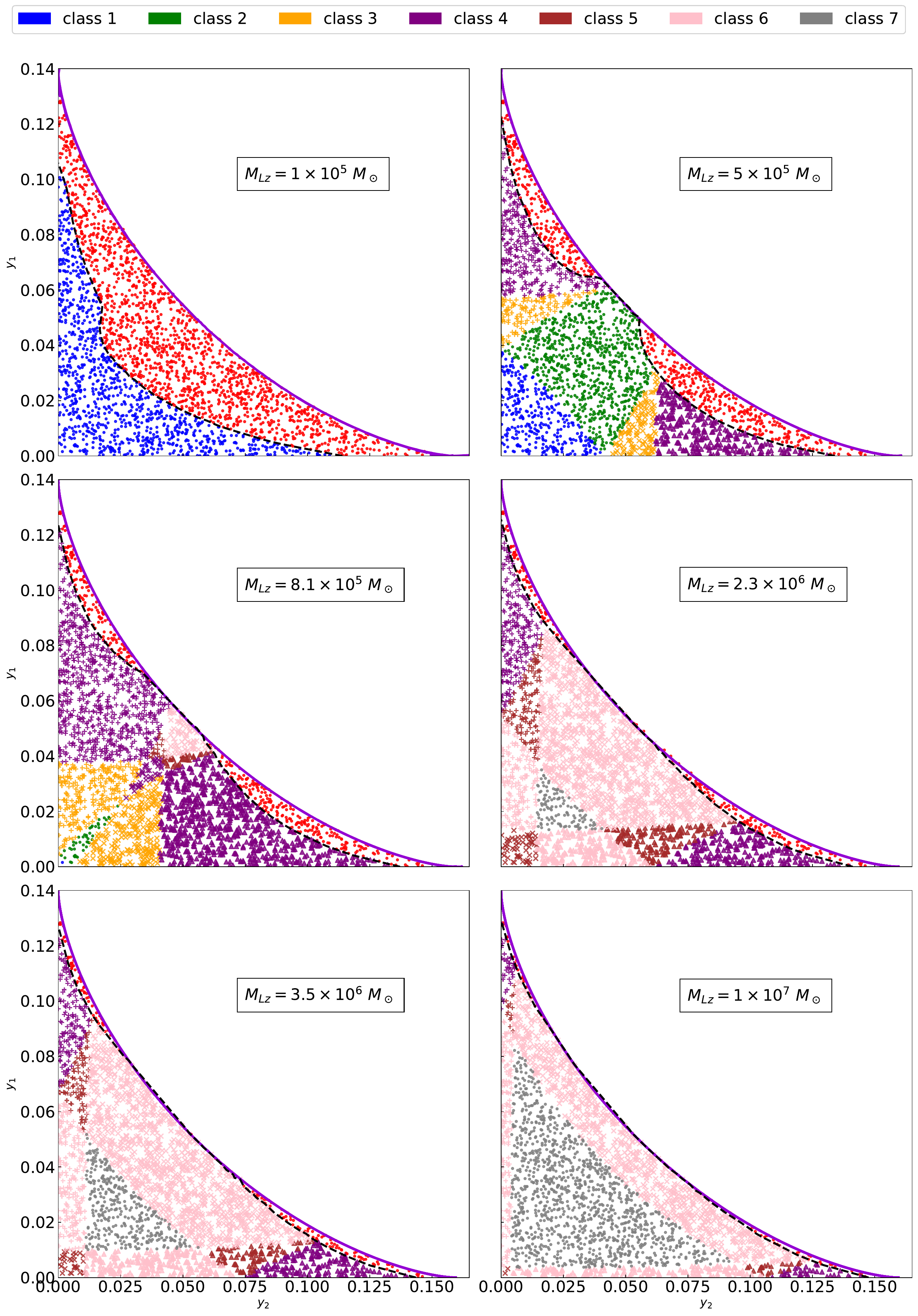}
        \caption{Image classification as a function of source position $\vec{y}$ for a single-isothermal ellipsoid (SIE) lens with $e = 0.2$ and redshifted lens mass $M_{Lz}$ increasing from $10^5 M_\odot$ (top left panel) to $10^7 M_\odot$ (bottom right panel). The geometrical-optics approximation fails for the red points, while colors indicating events of different classes are shown in the title box. Plus (+), cross ($\times$), and triangle ($\triangle$) markers are used for subclasses a, b, and c.}
        \label{fig:micro-macro-classification}
\end{figure*}

\begin{table*}
\[
\begin{tabular}{|c||c|c|c|c|c|c|c|c|c|}
\hline
 & \multicolumn{3}{c|}{0.1} & \multicolumn{3}{c|}{0.5} & \multicolumn{3}{c|}{0.81} \\
\hline 
& 2L & 3A & 3B & 2L & 3A & 3B & 2L & 3A & 3B \\
\hline \hline
1 & $0.225^{+0.297}_{-0.121}$ & $0.097^{+0.134}_{-0.05}$ & $0.122^{+0.161}_{-0.059}$ 
& $0.256^{+0.287}_{-0.239}$ & $0.106^{+0.12}_{-0.087}$ & $0.136^{+0.15}_{-0.116}$ 
& $0.261^{+0.424}_{-0.186}$ & $0.119^{+0.206}_{-0.093}$ & $0.146^{+0.205}_{-0.111}$ \\ \hline
2 & --- & --- & --- 
& $0.2^{+0.238}_{-0.107}$ & $0.08^{+0.105}_{-0.044}$ & $0.105^{+0.133}_{-0.057}$ 
& $0.268^{+0.292}_{-0.244}$ & $0.109^{+0.119}_{-0.1}$ & $0.139^{+0.15}_{-0.128}$ \\ \hline
3a & --- & --- & --- 
& $0.211^{+0.242}_{-0.155}$ & $0.105^{+0.122}_{-0.073}$ & $0.093^{+0.106}_{-0.076}$ 
& $0.243^{+0.275}_{-0.204}$ & $0.106^{+0.119}_{-0.087}$ & $0.119^{+0.138}_{-0.106}$ \\ \hline
3b & --- & --- & --- 
& $0.212^{+0.238}_{-0.161}$ & $0.068^{+0.08}_{-0.057}$ & $0.132^{+0.147}_{-0.098}$ 
& $0.243^{+0.273}_{-0.203}$ & $0.091^{+0.107}_{-0.08}$ & $0.135^{+0.149}_{-0.112}$ \\ \hline
4a & --- & --- & --- 
& $0.172^{+0.211}_{-0.113}$ & $0.1^{+0.123}_{-0.058}$ & $0.068^{+0.087}_{-0.038}$ 
& $0.178^{+0.23}_{-0.103}$ & $0.093^{+0.119}_{-0.051}$ & $0.077^{+0.104}_{-0.039}$ \\ \hline
4b & --- & --- & --- 
& --- & --- & --- 
& $0.196^{+0.217}_{-0.175}$ & $0.08^{+0.089}_{-0.072}$ & $0.105^{+0.115}_{-0.095}$ \\ \hline
4c & --- & --- & --- 
& $0.177^{+0.215}_{-0.143}$ & $0.05^{+0.064}_{-0.029}$ & $0.123^{+0.144}_{-0.098}$ 
& $0.18^{+0.228}_{-0.124}$ & $0.056^{+0.078}_{-0.031}$ & $0.117^{+0.143}_{-0.077}$ \\ \hline
5a & --- & --- & --- 
& --- & --- & --- 
& $0.16^{0.179}_{-0.113}$ & $0.068^{+0.075}_{-0.049}$ & $0.086^{+0.095}_{-0.06}$ \\ \hline
5b & --- & --- & --- 
& --- & --- & --- 
& --- & --- & --- \\ \hline
5c & --- & --- & --- 
& --- & --- & --- 
& $0.155^{+0.176}_{-0.096}$ & $0.062^{+0.071}_{-0.038}$ & $0.086^{+0.097}_{-0.056}$ \\ \hline
6a & --- & --- & ---
& --- & --- & --- 
& --- & --- & --- \\ \hline
6b & --- & --- & --- 
& --- & --- & --- 
& $0.111^{+0.164}_{-0.027}$ & $0.047^{+0.067}_{-0.012}$ & $0.063^{+0.09}_{-0.015}$ \\ \hline
6c & --- & --- & --- 
& --- & --- & --- 
& --- & --- & --- \\ \hline
7 & --- & --- & --- & --- & --- & --- & --- & --- & --- \\ \hline \hline
\text{all}  & $0.225^{+0.297}_{-0.121}$ & $0.097^{+0.134}_{-0.05}$ & $0.122^{+0.161}_{-0.059}$ & $0.2^{+0.267}_{-0.129}$ & $0.085^{+0.116}_{-0.04}$ & $0.11^{+0.143}_{-0.054}$ & $0.193^{+0.264}_{-0.11}$ & $0.081^{+0.116}_{-0.035}$ & $0.106^{+0.143}_{-0.049}$ \\ \hline 
\end{tabular}
\]

\[
\begin{tabular}{|c||c|c|c|c|c|c|c|c|c|}
\hline
 & \multicolumn{3}{c|}{2.3} & \multicolumn{3}{c|}{3.5} & \multicolumn{3}{c|}{10} \\
\hline \hline
& 2L & 3A & 3B & 2L & 3A & 3B & 2L & 3A & 3B \\
\hline
1 & --- & --- & --- & --- & --- & --- & --- & --- & --- \\ \hline
2 & --- & --- & --- & --- & --- & --- & --- & --- & --- \\ \hline
3a & --- & --- & --- & --- & --- & --- & --- & --- & --- \\ \hline
3b & --- & --- & --- & --- & --- & --- & --- & --- & --- \\ \hline
4a & $0.168^{+0.212}_{-0.111}$ & $0.107^{+0.123}_{-0.069}$ & $0.056^{+0.083}_{-0.038}$ 
& $0.161^{+0.202}_{-0.11}$ & $0.108^{+0.124}_{-0.072}$ & $0.05^{+0.072}_{-0.026}$ 
& $0.141^{+0.172}_{-0.092}$ & $0.11^{+0.125}_{-0.069}$ & $0.027^{+0.043}_{-0.016}$ \\ \hline
4b & --- & --- & --- & --- & --- & --- & --- & --- & --- \\ \hline
4c & $0.166^{+0.21}_{-0.119}$ & $0.038^{+0.058}_{-0.02}$ & $0.124^{+0.143}_{-0.09}$ 
& $0.157^{+0.119}_{-0.112}$ & $0.033^{+0.05}_{-0.017}$ & $0.122^{+0.143}_{-0.086}$ 
& $0.14^{+0.17}_{-0.099}$ & $0.018^{+0.028}_{-0.009}$ & $0.122^{+0.139}_{-0.084}$ \\ \hline
5a & $0.207^{+0.225}_{-0.148}$ & $0.104^{+0.115}_{-0.082}$ & $0.088^{+0.103}_{-0.059}$ 
& $0.194^{+0.211}_{-0.141}$ & $0.105^{+0.177}_{-0.085}$ & $0.076^{+0.09}_{-0.052}$ 
& $0.164^{+0.178}_{-0.128}$ & $0.108^{+0.122}_{-0.09}$ & $0.047^{+0.055}_{-0.034}$ \\ \hline
5b & $0.275^{+0.293}_{-0.265}$ & $0.113^{+0.12}_{-0.106}$ & $0.142^{+0.15}_{-0.136}$ 
& $0.28^{+0.298}_{-0.271}$ & $0.115^{+0.122}_{-0.11}$ & $0.144^{+0.153}_{-0.145}$ 
& $0.288^{+0.301}_{-0.285}$ & $0.117^{+0.122}_{-0.09}$ & $0.147^{+0.361}_{-0.342}$ \\ \hline
5c & $0.205^{+0.224}_{-0.151}$ & $0.064^{+0.076}_{-0.044}$ & $0.129^{+0.141}_{-0.103}$ 
& $0.192^{+0.209}_{-0.137}$ & $0.053^{+0.064}_{-0.035}$ & $0.128^{+0.14}_{-0.099}$ 
& $0.16^{+0.175}_{-0.121}$ & $0.031^{+0.037}_{-0.031}$ & $0.124^{+0.136}_{-0.097}$ \\ \hline
6a & $0.248^{+0.268}_{-0.229}$ & $0.113^{+0.12}_{-0.105}$ & $0.118^{+0.134}_{-0.1}$ 
& $0.244^{+0.273}_{-0.214}$ & $0.114^{+0.12}_{-0.107}$ & $0.114^{+0.137}_{-0.088}$ 
& $0.226^{+0.281}_{-0.181}$ & $0.118^{+0.123}_{-0.112}$ & $0.093^{+0.141}_{-0.057}$ \\ \hline
6b & $0.164^{+0.219}_{-0.079}$ & $0.067^{+0.095}_{-0.032}$ & $0.089^{+0.121}_{-0.041}$ 
& $0.156^{+0.205}_{-0.072}$ & $0.063^{+0.093}_{-0.028}$ & $0.084^{+0.118}_{-0.037}$ 
& $0.126^{+0.163}_{-0.06}$ & $0.047^{+0.088}_{-0.02}$ & $0.065^{+0.107}_{-0.028}$ \\ \hline
6c & $0.247^{+0.267}_{-0.227}$ & $0.09^{+0.104}_{-0.073}$ & $0.141^{+0.148}_{-0.133}$ 
& $0.244^{0.272}_{-0.213}$ & $0.086^{+0.107}_{-0.064}$ & $0.142^{+0.148}_{-0.134}$ 
& $0.224^{+0.28}_{-0.18}$ & $0.068^{+0.111}_{-0.039}$ & $0.144^{+0.148}_{-0.132}$ \\ \hline
7 & $0.237^{+0.252}_{-0.227}$ & $0.098^{+0.104}_{-0.087}$ & $0.126^{+0.133}_{-0.115}$ 
& $0.229^{+0.255}_{-0.212}$ & $0.096^{+0.107}_{-0.077}$ & $0.123^{+0.136}_{-0.103}$ 
& $0.211^{+0.263}_{-0.17}$ & $0.089^{+0.112}_{-0.056}$ & $0.115^{+0.141}_{-0.078}$ \\ \hline \hline
\text{all} & $0.189^{+0.264}_{-0.1}$ & $0.079^{+0.116}_{-0.031}$ & $0.103^{+0.143}_{-0.043}$ & $0.188^{+0.263}_{-0.093}$ & $0.078^{+0.116}_{-0.029}$ & $0.102^{+0.143}_{-0.041}$ & $0.186^{+0.263}_{-0.083}$ & $0.077^{+0.117}_{-0.025}$ & $0.101^{+0.143}_{-0.036}$ \\ \hline
\end{tabular}
\]
\caption{Medians of the mismatch distributions of the 14 classes for SIE lenses with $e = 0.2$ with source positions inside its tangential caustic. The last row, ``all'', provides the combined mismatches of all classes. Errors refer to the 5th and 95th percentile, respectively. Each column corresponds to the lens masses as indicated in units of $10^6 M_\odot$. The sub-columns inside the lens masses represent two-image (2L), three-image (two minima and one saddle, 3A), and three-image (one minima and two saddle, 3B) lensed templates, in that order.}
\label{tab:class mismatches}
\end{table*}

\begin{figure*}[t!]
        \includegraphics[width=1\textwidth, keepaspectratio]{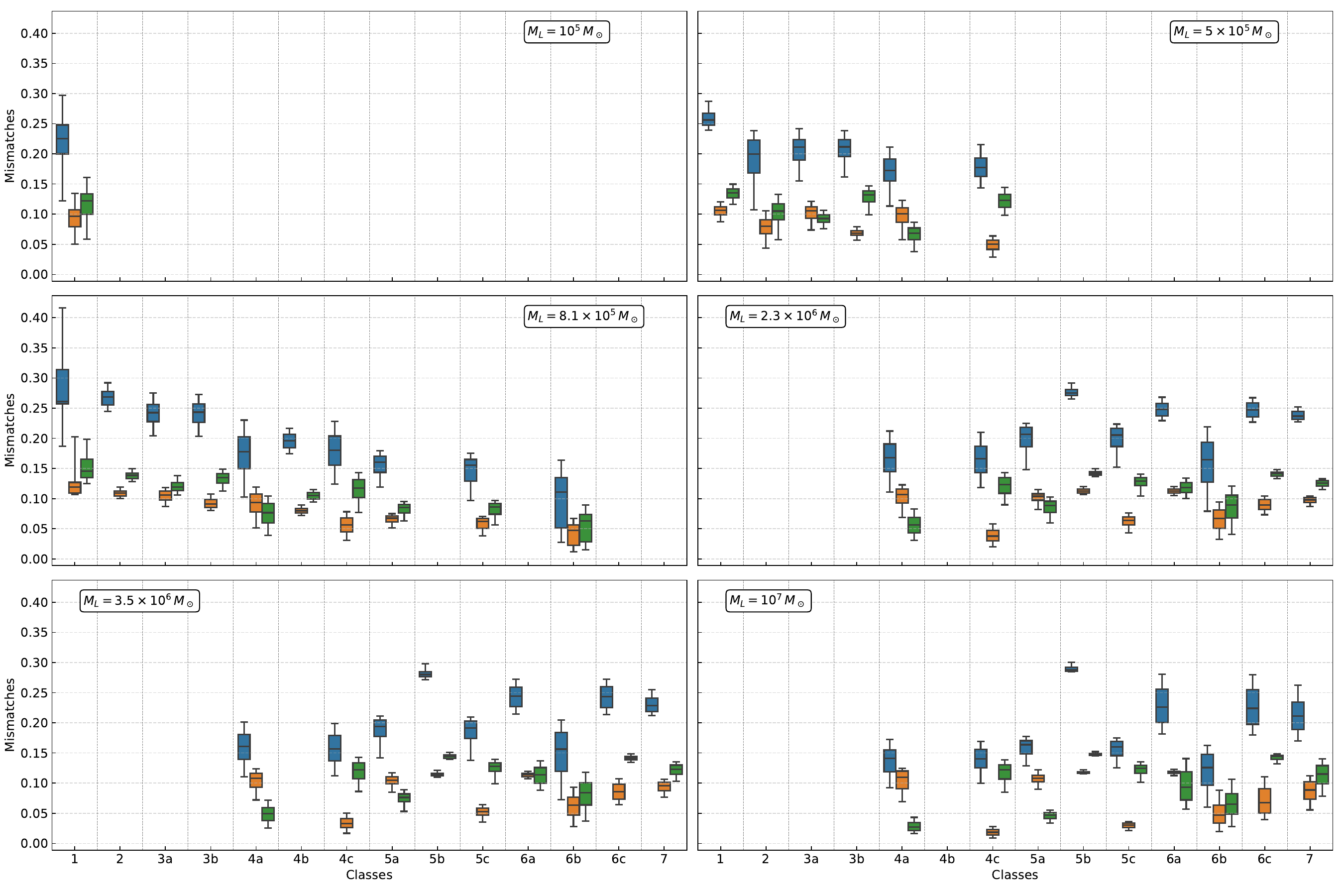}
        \caption{Mismatch between four-image lensed source and templates for different event classes across six different lens masses. Each box shows a median with edges representing the 25th and 75th percentiles. The whiskers extend to the 5th and 95th percentiles. The blue, orange, and green boxes represent two-image (2L), three-image (two minima and one saddle, 3A), and three-image (one minima and two saddles, 3B) lensed templates.}
        \label{fig:cls-distribution}
\end{figure*}

We plot $q_{{\rm max}}^{(j)}$ as a function of $y$ in Fig.~\ref{fig:q max} for $\phi = 0^\circ$, $45^\circ$, and $90^\circ$, $M_{z_S} = 50\,M_\odot$, and $M_{Lz} = 10^6 M_\odot$ and $10^8 M_\odot$.  The shape of these functions can be understood from the dependence of $\Delta t_d^{(j)}$ on $y$ shown in the bottom right panel of Fig.~\ref{fig:image param vs y images} and discussed in Sec.~\ref{sec:Singular Isothermal Ellipsoid (SIE)}.  Symmetry implies $\Delta t_d^{(2)} = 0$ ($\Delta t_d^{(3)} = \Delta t_d^{(4)}$) for $\phi = 0^\circ\,(90^\circ)$, so the curves corresponding to the second (fourth) image are absent in the left (right) panel.  In the limit $y \to y_t$, $\Delta t_d^{(3)} \to 0$ for $\phi = 0^\circ$ implying that $q_{{\rm max}}^{(3)} \to 1$.  For $\phi \neq 0^\circ$, $\Delta t_d^{(2)} \to \Delta t_d^{(3)}$ implying $q_{{\rm max}}^{(2)} \to q_{{\rm max}}^{(3)}$ in this limit.  In the opposite limit $y \to 0$, the lensing configuration approaches the Einstein cross with $\Delta t_d^{(2)} \to 0$ and $\Delta t_d^{(3)} \to \Delta t_d^{(4)}$ for all values of $\phi$ implying $q_{{\rm max}}^{(2)} \to 1$ and $q_{{\rm max}}^{(3)} \to q_{{\rm max}}^{(4)}$.  Since $\tau \propto \eta^{-1}$ and $\Delta t_d^{(j)} \propto M_{Lz}$, we see that $q_{{\rm max}}^{(j)} \propto M_{Lz}^{-1}$ for $q_{{\rm max}}^{(j)} \ll 1$.  For $M_{Lz} = 10^6 M_\odot$, $q_{{\rm max}}^{(j)} \gtrsim 0.1$ implying that comparable-mass BBH mergers similar to those observed by the LVK network are susceptible to microlensing for redshifted lens masses below this value.  However, for $M_{Lz} = 10^8 M_\odot$, $q_{{\rm max}}^{(j)} \lesssim 0.01$ implying that nearly all comparable-mass BBH mergers can only be macrolensed.

A given pair of images $j$ and $k$ is microlensed if
\begin{equation}
\Delta t_d^{(jk)} \equiv \Delta t_d^{(k)} - \Delta t_d^{(j)} < \tau\,.
\end{equation}
For lensed GW events with four images like the SIE model for $y < y_t$, this criterion allows us to categorize the events into 14 classes listed in Table~\ref{tab:classes}.  These classes only apply in the GO limit $M_{Lz} > M_{Lz ({\rm min})}^{(j)}$ for all $j$ given by Eq.~(\ref{eq:Mlz min}) which is needed for discrete images.  As general relativity is a scale-free theory, $\Delta t_d^{(j)} \propto M_{Lz}$ for a fixed lens model and dimensionless source position $\vec{y}$.  This implies that as the lens mass $M_{Lz}$ increases, a lensed GW event with fixed $\vec{y}$ will transition from the wave-optics to GO regime, then to increasing class number until reaching four macrolensed single images, i.e. class 7.  As the four images are sequentially ordered, only certain class transitions are possible; allowed transitions are listed in the eighth column of Table~\ref{tab:classes}.

In Fig.~\ref{fig:micro-macro-classification}, we classify lensed GW events as a function of source position $\vec{y}$ in the four-image regime for a SIE lens with $e = 0.2$ and redshifted lens masses $M_{Lz}$ between $10^5 M_\odot$ and $10^7 M_\odot$.  For a volume-limited survey, as will be expected for future observations of BBHs by third-generation ground-based GW detectors, sources will be evenly distributed in the lens plane, making the fraction of each image class proportional to their area in the source plane.  We provide these fractions in Table~\ref{tab:classes}.

For lenses with $M_{Lz} < M_{Lz({\rm min})}$ for all $y < y_t(\phi)$ ($M_{Lz} \lesssim 10^4 M_\odot$ for $e = 0.2$ as seen in Fig.~\ref{fig:deltaj ML four_imgs}), the GO approximation breaks down for all four-imaged lensed events.  As $M_{Lz}$ increases above this value, events with source positions near the origin (close to the Einstein cross configuration) begin falling into class 1 (four microlensed quadruple image) and only events closer to the tangential caustic at $y = y_t(\phi)$ remain in the wave-optics regime.  These wave-optics events constitute 59\% of events for $M_{Lz} = 10^5 M_\odot$ as shown in the top left panel of Fig.~\ref{fig:micro-macro-classification}, but this percentage monotonically decreases with lens mass and drops to only 2.3\% for $M_{Lz} = 10^7 M_\odot$ as shown in the bottom right panel.

As $M_{Lz}$ increases from $10^5 M_\odot$ to $5 \times 10^5 M_\odot$ as seen in the top two panels of Fig.~\ref{fig:micro-macro-classification}, events sufficiently far from the origin and caustic begin falling into class 2. For events closer to the major (minor) axis, the single image is the fourth (first) image, as can be inferred by the time delays shown in the bottom right panel of Fig.~\ref{fig:image param vs y images}.  As $M_{Lz}$ increases further, the area of class 2 in the source plane initially increases as it pushes towards the origin and caustic, then decreases as higher classes displace it.  

As shown in the top right panel of Fig.~\ref{fig:micro-macro-classification}, class 4 sources are notably concentrated near the cusp regions, a pattern that consistently appears across all panels. In these regions, sources positioned near the cusps produce three closely spaced images with nearly identical time delays, leading to prominent microlensing effects. Consequently, the area surrounding the cusp regions is predominantly occupied by class 4 sources, comprising approximately 24\% of all events, while the center of the plane is mainly populated by class 2 sources, representing around 31\%. As the source gradually moves away from the cusp toward the central region, the previously clustered images disperse and progressively form an Einstein ring configuration. This results in a transition from events of class 4 to those of class 3 which constitute $\sim 10\%$ of events for $M_{Lz} = 5 \times 10^5 M_\odot$.

With an increase in $M_{Lz}$ to $8.1 \times 10^5 M_\odot$, class 6b emerges near the fold, displacing class 2 toward the origin of the source plane. In class 6, only one pair of images undergoes microlensing. For sources positioned near the caustics of the relatively more massive lens (compared to $M_{Lz} = 5 \times 10^5 M_\odot$), the time delays become sufficiently large such that only two images appearing together near the critical curve exhibit small time delays, thus forming a microlensed image pair. In contrast, sources located closer to the center fall under classes 5a and 5c, where two distinct pairs of images experience microlensing. As illustrated in the middle panel, further increasing $M_{Lz}$ by a factor of approximately 2.8 causes class 6 to dominate, accounting for around 55\% of events. Meanwhile, the number of events in class 2 has reduced from approximately 2\% to 0\%. Sources near the origin now belong to class 5b, while a small region (4.6\%) in the central area of the plane is occupied by class 7. At the highest lens masses, class 4 is concentrated near the two cusps, where three images approach infinite magnification and zero time delay between each other.

In the bottom two panels of Fig.~\ref{fig:micro-macro-classification}, as \( M_{Lz} \) increases from \( 3.5 \times 10^6 M_\odot \) to \( 10^7 M_\odot \), classes 6 and 7 occupy an increasingly larger area of the source plane, simultaneously reducing the regions occupied by other classes such as class 4 and class 5. Specifically, the number of events in class 4 decreases from 13\% to 3.1\%, and in class 5 from 8.9\% to 2.2\%. For the highest lens mass considered, \( 10^7 M_\odot \), a dominant 92\% of the sources fall into classes 6 and 7. In the limit of large lens masses, the fraction of class 7 approaches unity.  Residual contributions of classes 4c and 5c exist near the major cusp, classes 4a and 5a near the minor cusp, class 6b near the fold, class 6c near the major axis, class 6a near the minor axis, and class 5b near the origin.

\section{\label{sec:Waveform mismatch: Unlensed and lensed templates}Waveform Mismatch: Lensed Templates}

The mismatch \(\epsilon\) between two GW waveforms provides a quantitative measure of how well a template waveform represents the true signal. For waveforms $h_1$ and $h_2$, \(\epsilon\) is defined as \cite{PhysRevD.49.2658}
\begin{equation}\label{eq:mismatch}
    \epsilon (h_1, h_2) \equiv 1 -  \max\limits_{t_c, \phi_c} \mathcal{O}(h_1, h_2),
\end{equation}
where the overlap \(\mathcal{O}(h_1, h_2)\) is given by:
\begin{equation}\label{eq:overlap}
    \mathcal{O}(h_1, h_2) = \frac{\langle h_1|h_2 \rangle}{\sqrt{\langle h_1|h_1 \rangle \langle h_2|h_2 \rangle}}.
\end{equation}
The inner product \(\langle h_1|h_2 \rangle\) is defined as:
\begin{equation}\label{eq:inner product}
    \langle h_1|h_2 \rangle = 4\,\text{Re} \int_{f_{\rm min}}^{f_{\rm cut}} df \frac{h_1^*(f) h_2(f)}{S_n(f)},
\end{equation}
where \(S_n(f)\) is the noise power spectral density. We compute \(\epsilon\) using the \textsc{pycbc.filter} package \cite{alex_nitz_2022_6324278}, generating GW signals from typical binary black hole (BBH) mergers observed by the LVK network, with mass ratio \(\eta = 0.25\) and chirp mass \(\mathcal{M} = 21 M_\odot\) (total mass \(M_{zS} = 50 M_\odot\)).

We evaluate the mismatch between simulated lensed GW signals and various template waveform families. Specifically, we consider templates corresponding to configurations producing two lensed images (hereafter 2L) - a minimum and a saddle point, three images arising from two minima and one saddle point (hereafter 3A), and three images from one minimum and two saddle points (hereafter 3B) of the Fermat potential. For the analysis, we assume the mismatch is minimized by parameterizing the templates using the properties of the brightest corresponding images within the true four-image signal.

Table~\ref{tab:class mismatches} summarizes the median mismatch values obtained for 14 distinct event classes across a range of lens masses. A consistent trend emerges across all investigated lens masses and event classes: the three-image lensed templates (3A and 3B) systematically yield lower mismatches compared to the two-image (2L) templates by a factor ranging from 1.5 to 5. All three template families improve in performance as the lens mass increases from $10^5 M_\odot$ to $10^7 M_\odot$.  A possible explanation for this improvement is the inclusion of source positions closer to the astroid at higher lens masses (fewer red points in Fig.~\ref{fig:micro-macro-classification}).  Near the fold (cusps), two (three) images dominate implying that template family 2L (3A or 3B) will be particularly effective.

The observed trends in mismatch values of the different templates agree with analytical estimates of the overlap integral derived under the extreme GO approximation, as detailed in Appendix~\ref{appendix:mismatch approximation}. Within this framework, incorporating information from additional lensed images, represented by their respective (positive definite) flux contributions, effectively incorporates more information about the source parameters, leading to improved waveform matches (i.e., higher overlap integrals). Template family 2L neglects both the first and fourth images, while template family 3A (3B) neglects only the fourth (first) image.  The relative performance of the two three-image template families will thus depend on whether the first or fourth image is brighter, i.e. whether the flux ratio $I^{(4)}$ is greater or less than unity.  If $I^{(4)} < 1$, template family 3A should perform better than template family 3B, while if $I^{(4)} > 1$, template family 3B should be superior.  The bottom left panel of Fig.~\ref{fig:image param vs y images} shows that $I^{(4)} < 1$ for source positions near the major axis where most of the area inside the asteroid is concentrated.
Consequently, the expected ordering of the median mismatches for the different template families is
\begin{equation}
\epsilon^{3A} < \epsilon^{3B} < \epsilon^{2L}
\end{equation}
as is seen to hold for all lens masses as shown in the bottom row of Table~\ref{tab:class mismatches}.

The 2L template family, constructed using only the two brightest lensed images (one minimum and one saddle-point image), lacks the information necessary to effectively distinguish between source locations near the major versus the minor cusp. As a result, it often yields comparable mismatch values for events situated symmetrically with respect to the lens center near these different cusp structures. For example, considering a lens mass $M_{Lz} = 2.3 \times 10^6 M_\odot$, the mismatch values obtained using the 2L template for event class 4a (associated with the minor cusp) and class 4c (associated with the major cusp) are $\epsilon^{2L} = 0.168$ and $\epsilon^{2L} = 0.166$, respectively, which are nearly identical. A similar proximity in mismatch values is observed for classes 5a ($\epsilon^{2L} = 0.207$) and 5c ($\epsilon^{2L} = 0.205$) under the same lensing configuration. 

The incorporation of a third lensed image, inherent in the 3A and 3B template families, breaks this degeneracy. This not only leads to generally lower mismatches, but also enables the differentiation between signals originating from proximity to the major versus the minor cusps. Template family 3A generally provides the best match for source positions near the major axis corresponding to image classes 3b, 4c, 5c, and 6c as shown in Fig.~\ref{fig:micro-macro-classification}.  Table~\ref{tab:class mismatches} shows median mismatches $\epsilon^{3A} < 0.1$ for these classes for all lens masses.  Template family 3B should perform comparatively better for source positions near the minor axis where $I^{(4)} > 1$.  This corresponds to image classes 3a, 4a, 5a, and 6a.  Table~\ref{tab:class mismatches} demonstrates this result, with template family 3B performing particularly well for image class 4a, corresponding to source positions near the minor cusp.  All three template families perform well for image class 6b (source positions near the fold) for which the second and third images are dominant.

These findings offer quantitative insights relevant to the construction of template banks for lensed GW searches. They underscore the potential benefits of incorporating three-image waveform models (both 3A and 3B types) to improve the efficiency and fidelity of detecting strongly lensed gravitational-wave signals.

\section{Discussion}

The accurate detection and robust characterization of gravitational-wave (GW) lensing events provide critical insights into both astrophysics and cosmology. This paper investigates GW lensing by elliptical mass distributions, specifically using the singular isothermal ellipsoid (SIE) lens model. Unlike simpler, axisymmetric lens models such as the singular isothermal sphere (SIS), the SIE model generates up to four images, offering a more realistic representation of potential astrophysical lensing scenarios.

Our analysis leverages the quasi-geometrical optics (quasi-GO) approximation to determine the limits of validity for the commonly used geometrical optics (GO) approximation for various lens masses and source locations with respect to the caustics. We found that the GO approximation holds robustly for lens masses $M_{Lz} \gtrsim 10^5 M_\odot$, within the sensitivity band of current ground-based GW detectors like the LVK network. However, significant wave-optics (WO) effects arise at lower lens masses or when the GW source approaches lens caustics, necessitating a detailed WO treatment to accurately interpret such events.

We also present a detailed classification of lensed GW events, with classes assigned according to the number and nature of their images. Distinguishing between macrolensing (multiple magnified copies of the same image, detected separately in time) and microlensing (simultaneously observed images due to overlapping signals within the detector's frequency band), our findings highlight how microlensing becomes increasingly relevant for lens masses below approximately $10^7 M_\odot$. For heavier lenses, images typically manifest as clearly temporally separated macrolensed events, significantly simplifying their identification and analysis.

Central to GW data analysis is accurately matching observed signals with theoretical templates. We quantified waveform mismatches between simulated four-image lensed GWs and template families, including two-image (2L) and three-image templates (3A and 3B). Our findings consistently demonstrate that three-image templates markedly outperform traditional two-image templates. For example, at class 2 of intermediate lens masses ($M_{Lz} \sim 5 \times 10^5 M_\odot$), using a two-image template typically yields mismatches on the order of $\epsilon^{2L} \sim 0.2$, whereas employing three-image templates reduces this mismatch significantly to $\epsilon^{3A} \sim 0.08$. Notably, these findings depend on the class, thus underscoring the need for context-specific template choices to optimize detection efficiency.

Interestingly, we found scenarios where the 3B template—representing configurations with one minimum and two saddle points—can surpass the 3A template in accuracy, particularly near minor cusp regions. Such nuanced results emphasize the importance of a diverse and comprehensive template bank to cover a broader range of realistic astrophysical lensing scenarios. Our analytic approximations in the extreme GO regime (presented in Appendix~\ref{appendix:mismatch approximation}) lend further theoretical support to these numerical findings, clearly delineating how additional lensed image information systematically enhances waveform matching and reduces detection ambiguity.

As GW detector sensitivity continues to improve and the observation of lensing events becomes increasingly probable, incorporating higher-order lensing effects into data-analysis pipelines becomes not merely beneficial but essential. Our findings demonstrate that templates accounting for more than two images are crucial to properly capture and interpret the complexity introduced by realistic elliptical lenses. This is particularly true for future observatories like Cosmic Explorer, Einstein Telescope, and space-based interferometers, which will have significantly lower frequency thresholds, thereby amplifying WO effects and complicating the identification of lensed events. 

Future research should extend our approach to incorporate populations of lenses as in \cite{Oguri:2018muv} and additional astrophysical complexities such as external shear, lens substructures, and realistic source distributions. Such extensions will enhance the precision of gravitational-wave lensing models and refine cosmological constraints derived from gravitational-wave observations.

\section*{\label{sec:Acknowledgement}Acknowledgements}

This work is supported by National Science Foundation Grant No. PHY-2309320. The authors acknowledge the Texas Advanced Computing Center (TACC) at The University of Texas at Austin for providing HPC resources that have contributed to the research results reported within this paper \cite{10.1145/3093338.3093385}. URL: \href{http://www.tacc.utexas.edu }{http://www.tacc.utexas.edu}

\appendix

\section{\label{appendix:mismatch approximation}Mismatch approximation in the extreme geometrical-optics limit}

This appendix details the derivation of the approximate overlap between a four-image lensed gravitational wave (GW) signal and various template waveforms in the extreme geometrical-optics limit. The core assumption is that the time delays between the lensed images are significantly larger than the inverse of the lower frequency cutoff of the detector, causing oscillatory terms in the overlap integrals to average to zero.

The overlap $\mathcal{O}$ between a signal waveform $\tilde{h}^{S}(f)$ and a template $\tilde{h}^{T}(f)$ is defined as:
\begin{align}\label{app:eq:overlap}
    \mathcal{O} \equiv \frac{\langle\tilde{h}^{T}(f)| \tilde{h}^{S}(f)\rangle}{\sqrt{\langle\tilde{h}^{S}(f)|\tilde{h}^{S}(f)\rangle \langle\tilde{h}^T(f)|\tilde{h}^T(f)\rangle}} \,,
\end{align}
where the noise-weighted inner product is given by:
\begin{equation}\label{app:eq:inner_product}
    \langle h_1|h_2 \rangle = 4\,\text{Re} \int_{f_{\rm min}}^{f_{\rm cut}} df \frac{h_1^*(f) h_2(f)}{S_n(f)} \,,
\end{equation}
Here, $S_n(f)$ is the power spectral density (PSD) of the detector noise. A lensed waveform $\tilde{h}^{\text{L}}(f)$ is related to the unlensed waveform $\tilde{h}(f)$ via the lensing amplification factor, $F(f)$, as $\tilde{h}^{\text{L}}(f) = F(f) \tilde{h}(f)$.

The signal is assumed to be a four-image lensed GW, composed of two minima and two saddle points (images) of the Fermat potential. Its amplification factor $F^{4S}(f)$ is 
\begin{align} \label{app:eq:four_image_source}
    F^{4S}(f) &= |\mu^{(1)}|^{1/2}\Bigl[1 + \sqrt{I_2} e^{i 2\pi f \Delta t^{(21)}} \nonumber\\
    &\quad - i \sqrt{I_3} e^{i 2\pi f \Delta t^{(31)}} - i \sqrt{I_4} e^{i 2\pi f \Delta t^{(41)}}\Bigr] \,,
\end{align}
where $I_j \equiv |\mu^{(j)}|/|\mu^{(1)}|$ is the flux ratio of the $j$-th image relative to the image 1, and $\Delta t^{(ij)} = t^{(j)} - t^{(i)}$ is the time delay between images $i$ and $j$ given by Eq.~(\ref{eq:dimensionless time delay}). The factors of $-i$ for images 3 and 4 account for the $\pi/2$ Morse phase accumulated by saddle-point images.

The template bank consists of waveforms corresponding to two or three lensed images. A two-image template (template 2L), including the brightest minimum and the brightest saddle point, is given by
\begin{align} \label{app:eq:two-image_template}
    F^{2L}(f) = |\mu^{(1)}|^{1/2}\left(\sqrt{I_2} e^{i 2\pi f \Delta t^{(21)}} - i\sqrt{I_3} e^{i 2\pi f \Delta t^{(31)}}\right)
\end{align}
A three-image template can include either both minima and the brightest saddle point (template 3A).
\begin{align}
    F^{3A}(f) = |\mu^{(1)}|^{1/2}\left[1 + \sqrt{I_2} e^{i 2\pi f \Delta t^{(21)}} - i\sqrt{I_3}e^{i 2\pi f \Delta t^{(31)}}\right] \label{app:eq:three-image_template_A}
\end{align}
or the brightest minimum and both saddle-points (template 3B).
\begin{align}
    F^{3B}(f) &= |\mu^{(1)}|^{1/2}\left[\sqrt{I_2} e^{i 2\pi f \Delta t^{(21)}} - i\sqrt{I_3} e^{i 2\pi f \Delta t^{(31)}} \right. \notag \\ & \left. - i \sqrt{I_4}e^{i 2\pi f \Delta t^{(41)}}\right] \label{app:eq:three-image_template_B}
\end{align}

To compute the overlap, we require the products $|F^{4S}|^2$, $|F^{T}|^2$, and $\text{Re}[F^{T*} F^{4S}]$. The signal norm can be written as
\begin{align}\label{app:eq:4s4s}
|F^{4S}|^2 / |\mu^{(1)}| &= 1 + I_2 + I_3 + I_4 + 2\sqrt{I_2}\cos(2\pi f \Delta t^{(21)}) \nonumber \\
&\quad + 2\sqrt{I_3}\cos(2\pi f \Delta t^{(31)}) \nonumber \\
&\quad + 2\sqrt{I_4}\cos(2\pi f \Delta t^{(41)}) \nonumber \\
&\quad + 2\sqrt{I_2I_3}\cos(2\pi f \Delta t^{(32)}) \nonumber \\
&\quad + 2\sqrt{I_2I_4}\cos(2\pi f \Delta t^{(42)}) \nonumber \\
&\quad + 2\sqrt{I_3I_4}\cos(2\pi f \Delta t^{(43)}) \,,
\end{align}
and the template norms are 
\begin{align}
|F^{2L}|^2 / |\mu^{(1)}| &= I_2 + I_3 + 2 \sqrt{I_2I_3}\sin(2\pi f \Delta t^{(32)}) \label{app:eq:2l2l} \\
|F^{3A}|^2 / |\mu^{(1)}| &= 1 + I_2 + I_3 + 2\sqrt{I_2}\cos(2\pi f \Delta t^{(21)}) \nonumber \\
&\quad + 2\sqrt{I_3}\sin(2\pi f \Delta t^{(31)}) \nonumber \\ &\quad + 2\sqrt{I_2I_3}\sin(2\pi f \Delta t^{(32)}) \label{app:eq:3a3a} \\
|F^{3B}|^2 / |\mu^{(1)}| &= I_2 + I_3 + I_4 + 2\sqrt{I_2I_3} \sin(2\pi f \Delta t^{(32)}) \nonumber \\
&\quad + 2\sqrt{I_2I_4} \sin(2\pi f \Delta t^{(42)}) \nonumber \\
&\quad + 2\sqrt{I_3I_4} \cos(2\pi f \Delta t^{(43)}) \,. \label{app:eq:3b3b}
\end{align}

The cross-products are given by
\begin{align}
\text{Re}[F^{2L*}F^{4S}] / |\mu^{(1)}| &= I_2 + I_3 + \sqrt{I_2}\cos(2\pi f \Delta t^{(21)}) \nonumber \\
&\quad + \sqrt{I_3}\sin(2\pi f     \Delta t^{(31)})  \nonumber \\
&\quad + 2\sqrt{I_2I_3}\sin(2\pi f \Delta t^{(32)}) \nonumber \\
&\quad + \sqrt{I_2I_4}\sin(2\pi f  \Delta t^{(42)}) \nonumber \\
&\quad + \sqrt{I_3I_4}\cos(2\pi f  \Delta t^{(43)}) \label{app:eq:2l4s} \\
\text{Re}[F^{3A*}F^{4S}] / |\mu^{(1)}| &= 1 + I_2 + I_3 + 2\sqrt{I_2}\cos(2\pi f             \Delta t^{(21)}) \nonumber \\
&\quad + 2\sqrt{I_3}\sin(2\pi f    \Delta t^{(31)}) \nonumber \\
&\quad + 2\sqrt{I_2I_3}\sin(2\pi f \Delta t^{(32)}) \nonumber \\
&\quad + \sqrt{I_4}\sin(2\pi f     \Delta t^{(41)}) \nonumber \\
&\quad + \sqrt{I_2I_4}\sin(2\pi f  \Delta t^{(42)}) \nonumber \\
&\quad + \sqrt{I_3I_4}\cos(2\pi f  \Delta t^{(43)}) \label{app:eq:3a4s} \\
\text{Re}[F^{3B*}F^{4S}] / |\mu^{(1)}| &= I_2 + I_3 + I_4 + \sqrt{I_2}\cos(2\pi f              \Delta t^{(21)}) \nonumber \\
&\quad + \sqrt{I_3}\sin(2\pi f     \Delta t^{(31)}) \nonumber \\
&\quad + \sqrt{I_4}\sin(2\pi f     \Delta t^{(41)}) \nonumber \\
&\quad + 2\sqrt{I_2I_3}\sin(2\pi f \Delta t^{(32)}) \nonumber \\
&\quad + 2\sqrt{I_2I_4}\sin(2\pi f \Delta t^{(42)}) \nonumber \\
&\quad+ 2\sqrt{I_3I_4}\cos(2\pi f  \Delta t^{(43)}) \,. \label{app:eq:3b4s}
\end{align}

We now consider the extreme geometrical-optics limit where the time delays are large, such that $2\pi f_{\rm min} \Delta t^{(ij)} \gg 1$ for all $i \neq j$. In this regime, the trigonometric terms $\cos(2\pi f \Delta t^{(ij)})$ and $\sin(2\pi f \Delta t^{(ij)})$ are highly oscillatory within the integration band. By the Riemann-Lebesgue lemma, their integrals over the band are suppressed and can be approximated as zero. Therefore, only the non-oscillatory (constant) terms in the products of amplification factors contribute to the inner product integrals. Let $\mathcal{N}^2 \equiv 4 \int_{f_{\text{min}}}^{f_{\text{cut}}} |\tilde{h}(f)|^2 / S_n(f) df$. The inner products simplify as follows:
\begin{align}
    \langle \tilde{h}^{S}|\tilde{h}^{S} \rangle &\approx \left[ |F^{S}|^2 \right]_{\text{non-osc}} \mathcal{N}^2 \nonumber \\
    &\quad= |\mu^{(1)}| (1+I_2+I_3+I_4) \mathcal{N}^2 \equiv |\mu^{(1)}| \mathcal{I}_{\text{tot}} \mathcal{N}^2 \\
    \langle \tilde{h}^{T}|\tilde{h}^{T} \rangle &\approx \left[ |F^{T}|^2 \right]_{\text{non-osc}} \mathcal{N}^2 \\
    \langle \tilde{h}^{T}|\tilde{h}^{S} \rangle &\approx \left[ \text{Re}[F^{T*}F^{S}] \right]_{\text{non-osc}} \mathcal{N}^2
\end{align}
where $\mathcal{I}_{\text{tot}} = 1+I_2+I_3+I_4$ is the total magnification.

Applying this approximation to our templates, for the template 2L, the non-oscillatory parts of Eqs.\,\eqref{app:eq:2l2l} and \eqref{app:eq:2l4s} are $(I_2+I_3)$. The overlap becomes
\begin{align} \label{app:eq:o2L}
    \mathcal{O}^{2L} &\to \frac{|\mu^{(1)}|(I_2+I_3)\mathcal{N}^2}{\sqrt{\left(|\mu^{(1)}|\mathcal{I}_{\text{tot}}\mathcal{N}^2\right) \left(|\mu^{(1)}|(I_2+I_3)\mathcal{N}^2\right)}} \nonumber \\
    &\quad= \sqrt{\frac{I_2+I_3}{\mathcal{I}_{\text{tot}}}}
\end{align}
For the template 3A, the non-oscillatory parts of Eqs.\,\eqref{app:eq:3a3a} and \eqref{app:eq:3a4s} are $(1+I_2+I_3)$. The overlap is:
\begin{align} \label{app:eq:overlap_3A}
    \mathcal{O}^{3A} \to \sqrt{\frac{1+I_2+I_3}{\mathcal{I}_{\text{tot}}}}
\end{align}
For the template 3B, the non-oscillatory parts of Eqs.\,\eqref{app:eq:3b3b} and \eqref{app:eq:3b4s} are $(I_2+I_3+I_4)$. The overlap is:
\begin{align} \label{app:eq:overlap_3B}
    \mathcal{O}^{3B} \to \sqrt{\frac{I_2+I_3+I_4}{\mathcal{I}_{\text{tot}}}}
\end{align}

\FloatBarrier

\bibliography{bibme}

\begin{thebibliography}{65}%
\makeatletter
\providecommand \@ifxundefined [1]{%
 \@ifx{#1\undefined}
}%
\providecommand \@ifnum [1]{%
 \ifnum #1\expandafter \@firstoftwo
 \else \expandafter \@secondoftwo
 \fi
}%
\providecommand \@ifx [1]{%
 \ifx #1\expandafter \@firstoftwo
 \else \expandafter \@secondoftwo
 \fi
}%
\providecommand \natexlab [1]{#1}%
\providecommand \enquote  [1]{``#1''}%
\providecommand \bibnamefont  [1]{#1}%
\providecommand \bibfnamefont [1]{#1}%
\providecommand \citenamefont [1]{#1}%
\providecommand \href@noop [0]{\@secondoftwo}%
\providecommand \href [0]{\begingroup \@sanitize@url \@href}%
\providecommand \@href[1]{\@@startlink{#1}\@@href}%
\providecommand \@@href[1]{\endgroup#1\@@endlink}%
\providecommand \@sanitize@url [0]{\catcode `\\12\catcode `\$12\catcode
  `\&12\catcode `\#12\catcode `\^12\catcode `\_12\catcode `\%12\relax}%
\providecommand \@@startlink[1]{}%
\providecommand \@@endlink[0]{}%
\providecommand \url  [0]{\begingroup\@sanitize@url \@url }%
\providecommand \@url [1]{\endgroup\@href {#1}{\urlprefix }}%
\providecommand \urlprefix  [0]{URL }%
\providecommand \Eprint [0]{\href }%
\providecommand \doibase [0]{https://doi.org/}%
\providecommand \selectlanguage [0]{\@gobble}%
\providecommand \bibinfo  [0]{\@secondoftwo}%
\providecommand \bibfield  [0]{\@secondoftwo}%
\providecommand \translation [1]{[#1]}%
\providecommand \BibitemOpen [0]{}%
\providecommand \bibitemStop [0]{}%
\providecommand \bibitemNoStop [0]{.\EOS\space}%
\providecommand \EOS [0]{\spacefactor3000\relax}%
\providecommand \BibitemShut  [1]{\csname bibitem#1\endcsname}%
\let\auto@bib@innerbib\@empty
\bibitem [{\citenamefont {{Einstein}}(1936)}]{1936Sci....84..506E}%
  \BibitemOpen
  \bibfield  {author} {\bibinfo {author} {\bibfnamefont {A.}~\bibnamefont
  {{Einstein}}},\ }\href {https://doi.org/10.1126/science.84.2188.506}
  {\bibfield  {journal} {\bibinfo  {journal} {Science}\ }\textbf {\bibinfo
  {volume} {84}},\ \bibinfo {pages} {506} (\bibinfo {year} {1936})}\BibitemShut
  {NoStop}%
\bibitem [{\citenamefont {{Dalal}}\ and\ \citenamefont
  {{Kochanek}}(2002)}]{2002ApJ...572...25D}%
  \BibitemOpen
  \bibfield  {author} {\bibinfo {author} {\bibfnamefont {N.}~\bibnamefont
  {{Dalal}}}\ and\ \bibinfo {author} {\bibfnamefont {C.~S.}\ \bibnamefont
  {{Kochanek}}},\ }\href {https://doi.org/10.1086/340303} {\bibfield  {journal}
  {\bibinfo  {journal} {\apj}\ }\textbf {\bibinfo {volume} {572}},\ \bibinfo
  {pages} {25} (\bibinfo {year} {2002})},\ \Eprint
  {https://arxiv.org/abs/astro-ph/0111456} {arXiv:astro-ph/0111456 [astro-ph]}
  \BibitemShut {NoStop}%
\bibitem [{\citenamefont {{Oguri}}(2019)}]{2019RPPh...82l6901O}%
  \BibitemOpen
  \bibfield  {author} {\bibinfo {author} {\bibfnamefont {M.}~\bibnamefont
  {{Oguri}}},\ }\href {https://doi.org/10.1088/1361-6633/ab4fc5} {\bibfield
  {journal} {\bibinfo  {journal} {Reports on Progress in Physics}\ }\textbf
  {\bibinfo {volume} {82}},\ \bibinfo {eid} {126901} (\bibinfo {year}
  {2019})},\ \Eprint {https://arxiv.org/abs/1907.06830} {arXiv:1907.06830
  [astro-ph.CO]} \BibitemShut {NoStop}%
\bibitem [{\citenamefont {{Treu}}\ and\ \citenamefont
  {{Marshall}}(2016)}]{2016A&ARv..24...11T}%
  \BibitemOpen
  \bibfield  {author} {\bibinfo {author} {\bibfnamefont {T.}~\bibnamefont
  {{Treu}}}\ and\ \bibinfo {author} {\bibfnamefont {P.~J.}\ \bibnamefont
  {{Marshall}}},\ }\href {https://doi.org/10.1007/s00159-016-0096-8} {\bibfield
   {journal} {\bibinfo  {journal} {\aapr}\ }\textbf {\bibinfo {volume} {24}},\
  \bibinfo {eid} {11} (\bibinfo {year} {2016})},\ \Eprint
  {https://arxiv.org/abs/1605.05333} {arXiv:1605.05333 [astro-ph.CO]}
  \BibitemShut {NoStop}%
\bibitem [{\citenamefont {{Bartelmann}}(2010)}]{2010CQGra..27w3001B}%
  \BibitemOpen
  \bibfield  {author} {\bibinfo {author} {\bibfnamefont {M.}~\bibnamefont
  {{Bartelmann}}},\ }\href {https://doi.org/10.1088/0264-9381/27/23/233001}
  {\bibfield  {journal} {\bibinfo  {journal} {Classical and Quantum Gravity}\
  }\textbf {\bibinfo {volume} {27}},\ \bibinfo {eid} {233001} (\bibinfo {year}
  {2010})},\ \Eprint {https://arxiv.org/abs/1010.3829} {arXiv:1010.3829
  [astro-ph.CO]} \BibitemShut {NoStop}%
\bibitem [{\citenamefont {{Brada{\v{c}}}}\ \emph {et~al.}(2002)\citenamefont
  {{Brada{\v{c}}}}, \citenamefont {{Schneider}}, \citenamefont {{Steinmetz}},
  \citenamefont {{Lombardi}}, \citenamefont {{King}},\ and\ \citenamefont
  {{Porcas}}}]{2002A&A...388..373B}%
  \BibitemOpen
  \bibfield  {author} {\bibinfo {author} {\bibfnamefont {M.}~\bibnamefont
  {{Brada{\v{c}}}}}, \bibinfo {author} {\bibfnamefont {P.}~\bibnamefont
  {{Schneider}}}, \bibinfo {author} {\bibfnamefont {M.}~\bibnamefont
  {{Steinmetz}}}, \bibinfo {author} {\bibfnamefont {M.}~\bibnamefont
  {{Lombardi}}}, \bibinfo {author} {\bibfnamefont {L.~J.}\ \bibnamefont
  {{King}}},\ and\ \bibinfo {author} {\bibfnamefont {R.}~\bibnamefont
  {{Porcas}}},\ }\href {https://doi.org/10.1051/0004-6361:20020559} {\bibfield
  {journal} {\bibinfo  {journal} {\aap}\ }\textbf {\bibinfo {volume} {388}},\
  \bibinfo {pages} {373} (\bibinfo {year} {2002})},\ \Eprint
  {https://arxiv.org/abs/astro-ph/0112038} {arXiv:astro-ph/0112038 [astro-ph]}
  \BibitemShut {NoStop}%
\bibitem [{\citenamefont {{Halkola}}\ \emph {et~al.}(2006)\citenamefont
  {{Halkola}}, \citenamefont {{Seitz}},\ and\ \citenamefont
  {{Pannella}}}]{2006MNRAS.372.1425H}%
  \BibitemOpen
  \bibfield  {author} {\bibinfo {author} {\bibfnamefont {A.}~\bibnamefont
  {{Halkola}}}, \bibinfo {author} {\bibfnamefont {S.}~\bibnamefont {{Seitz}}},\
  and\ \bibinfo {author} {\bibfnamefont {M.}~\bibnamefont {{Pannella}}},\
  }\href {https://doi.org/10.1111/j.1365-2966.2006.10948.x} {\bibfield
  {journal} {\bibinfo  {journal} {\mnras}\ }\textbf {\bibinfo {volume} {372}},\
  \bibinfo {pages} {1425} (\bibinfo {year} {2006})},\ \Eprint
  {https://arxiv.org/abs/astro-ph/0605470} {arXiv:astro-ph/0605470 [astro-ph]}
  \BibitemShut {NoStop}%
\bibitem [{\citenamefont {Abbott}\ \emph
  {et~al.}(2021{\natexlab{a}})\citenamefont {Abbott}, \citenamefont {Abbott},
  \citenamefont {Acernese}, \citenamefont {Ackley}, \citenamefont {Adams},
  \citenamefont {Adhikari}, \citenamefont {Adhikari}, \citenamefont {Adya},
  \citenamefont {Affeldt}, \citenamefont {Agarwal} \emph
  {et~al.}}]{abbott2021gwtc-3}%
  \BibitemOpen
  \bibfield  {author} {\bibinfo {author} {\bibfnamefont {R.}~\bibnamefont
  {Abbott}}, \bibinfo {author} {\bibfnamefont {T.}~\bibnamefont {Abbott}},
  \bibinfo {author} {\bibfnamefont {F.}~\bibnamefont {Acernese}}, \bibinfo
  {author} {\bibfnamefont {K.}~\bibnamefont {Ackley}}, \bibinfo {author}
  {\bibfnamefont {C.}~\bibnamefont {Adams}}, \bibinfo {author} {\bibfnamefont
  {N.}~\bibnamefont {Adhikari}}, \bibinfo {author} {\bibfnamefont
  {R.}~\bibnamefont {Adhikari}}, \bibinfo {author} {\bibfnamefont
  {V.}~\bibnamefont {Adya}}, \bibinfo {author} {\bibfnamefont {C.}~\bibnamefont
  {Affeldt}}, \bibinfo {author} {\bibfnamefont {D.}~\bibnamefont {Agarwal}},
  \emph {et~al.},\ }\href@noop {} {\bibfield  {journal} {\bibinfo  {journal}
  {arXiv preprint arXiv:2111.03606}\ } (\bibinfo {year}
  {2021}{\natexlab{a}})}\BibitemShut {NoStop}%
\bibitem [{\citenamefont {Abbott}\ \emph
  {et~al.}(2021{\natexlab{b}})\citenamefont {Abbott}, \citenamefont {Abe},
  \citenamefont {Acernese}, \citenamefont {Ackley}, \citenamefont {Adhikari},
  \citenamefont {Adhikari}, \citenamefont {Adkins}, \citenamefont {Adya},
  \citenamefont {Affeldt}, \citenamefont {Agarwal} \emph
  {et~al.}}]{abbott2021tests}%
  \BibitemOpen
  \bibfield  {author} {\bibinfo {author} {\bibfnamefont {R.}~\bibnamefont
  {Abbott}}, \bibinfo {author} {\bibfnamefont {H.}~\bibnamefont {Abe}},
  \bibinfo {author} {\bibfnamefont {F.}~\bibnamefont {Acernese}}, \bibinfo
  {author} {\bibfnamefont {K.}~\bibnamefont {Ackley}}, \bibinfo {author}
  {\bibfnamefont {N.}~\bibnamefont {Adhikari}}, \bibinfo {author}
  {\bibfnamefont {R.}~\bibnamefont {Adhikari}}, \bibinfo {author}
  {\bibfnamefont {V.}~\bibnamefont {Adkins}}, \bibinfo {author} {\bibfnamefont
  {V.}~\bibnamefont {Adya}}, \bibinfo {author} {\bibfnamefont {C.}~\bibnamefont
  {Affeldt}}, \bibinfo {author} {\bibfnamefont {D.}~\bibnamefont {Agarwal}},
  \emph {et~al.},\ }\href@noop {} {\bibfield  {journal} {\bibinfo  {journal}
  {arXiv preprint arXiv:2112.06861}\ } (\bibinfo {year}
  {2021}{\natexlab{b}})}\BibitemShut {NoStop}%
\bibitem [{\citenamefont {Abbott}\ \emph
  {et~al.}(2021{\natexlab{c}})\citenamefont {Abbott}, \citenamefont {Abbott},
  \citenamefont {Abraham}, \citenamefont {Acernese}, \citenamefont {Ackley},
  \citenamefont {Adams}, \citenamefont {Adams}, \citenamefont {Adhikari},
  \citenamefont {Adya}, \citenamefont {Affeldt},\ and\ \citenamefont
  {et~al.}}]{2021-pop}%
  \BibitemOpen
  \bibfield  {author} {\bibinfo {author} {\bibfnamefont {R.}~\bibnamefont
  {Abbott}}, \bibinfo {author} {\bibfnamefont {T.~D.}\ \bibnamefont {Abbott}},
  \bibinfo {author} {\bibfnamefont {S.}~\bibnamefont {Abraham}}, \bibinfo
  {author} {\bibfnamefont {F.}~\bibnamefont {Acernese}}, \bibinfo {author}
  {\bibfnamefont {K.}~\bibnamefont {Ackley}}, \bibinfo {author} {\bibfnamefont
  {A.}~\bibnamefont {Adams}}, \bibinfo {author} {\bibfnamefont
  {C.}~\bibnamefont {Adams}}, \bibinfo {author} {\bibfnamefont {R.~X.}\
  \bibnamefont {Adhikari}}, \bibinfo {author} {\bibfnamefont {V.~B.}\
  \bibnamefont {Adya}}, \bibinfo {author} {\bibfnamefont {C.}~\bibnamefont
  {Affeldt}},\ and\ \bibinfo {author} {\bibnamefont {et~al.}},\ }\href
  {https://doi.org/10.3847/2041-8213/abe949} {\bibfield  {journal} {\bibinfo
  {journal} {The Astrophysical Journal Letters}\ }\textbf {\bibinfo {volume}
  {913}},\ \bibinfo {pages} {L7} (\bibinfo {year}
  {2021}{\natexlab{c}})}\BibitemShut {NoStop}%
\bibitem [{\citenamefont {Abbott}\ \emph {et~al.}(2018)\citenamefont {Abbott},
  \citenamefont {Abbott}, \citenamefont {Abbott}, \citenamefont {Acernese},
  \citenamefont {Ackley}, \citenamefont {Adams}, \citenamefont {Adams},
  \citenamefont {Addesso}, \citenamefont {Adhikari}, \citenamefont {Adya} \emph
  {et~al.}}]{abbott2018gw170817}%
  \BibitemOpen
  \bibfield  {author} {\bibinfo {author} {\bibfnamefont {B.~P.}\ \bibnamefont
  {Abbott}}, \bibinfo {author} {\bibfnamefont {R.}~\bibnamefont {Abbott}},
  \bibinfo {author} {\bibfnamefont {T.}~\bibnamefont {Abbott}}, \bibinfo
  {author} {\bibfnamefont {F.}~\bibnamefont {Acernese}}, \bibinfo {author}
  {\bibfnamefont {K.}~\bibnamefont {Ackley}}, \bibinfo {author} {\bibfnamefont
  {C.}~\bibnamefont {Adams}}, \bibinfo {author} {\bibfnamefont
  {T.}~\bibnamefont {Adams}}, \bibinfo {author} {\bibfnamefont
  {P.}~\bibnamefont {Addesso}}, \bibinfo {author} {\bibfnamefont {R.~X.}\
  \bibnamefont {Adhikari}}, \bibinfo {author} {\bibfnamefont {V.~B.}\
  \bibnamefont {Adya}}, \emph {et~al.},\ }\href@noop {} {\bibfield  {journal}
  {\bibinfo  {journal} {Physical review letters}\ }\textbf {\bibinfo {volume}
  {121}},\ \bibinfo {pages} {161101} (\bibinfo {year} {2018})}\BibitemShut
  {NoStop}%
\bibitem [{\citenamefont {{Lawrence}}(1971)}]{1971NCimB...6..225L}%
  \BibitemOpen
  \bibfield  {author} {\bibinfo {author} {\bibfnamefont {J.~K.}\ \bibnamefont
  {{Lawrence}}},\ }\href {https://doi.org/10.1007/BF02735388} {\bibfield
  {journal} {\bibinfo  {journal} {Nuovo Cimento B Serie}\ }\textbf {\bibinfo
  {volume} {6B}},\ \bibinfo {pages} {225} (\bibinfo {year} {1971})}\BibitemShut
  {NoStop}%
\bibitem [{\citenamefont {{Ohanian}}(1974)}]{1974IJTP....9..425O}%
  \BibitemOpen
  \bibfield  {author} {\bibinfo {author} {\bibfnamefont {H.~C.}\ \bibnamefont
  {{Ohanian}}},\ }\href {https://doi.org/10.1007/BF01810927} {\bibfield
  {journal} {\bibinfo  {journal} {International Journal of Theoretical
  Physics}\ }\textbf {\bibinfo {volume} {9}},\ \bibinfo {pages} {425} (\bibinfo
  {year} {1974})}\BibitemShut {NoStop}%
\bibitem [{\citenamefont {{Nakamura}}\ and\ \citenamefont
  {{Deguchi}}(1999)}]{1999PThPS.133..137N}%
  \BibitemOpen
  \bibfield  {author} {\bibinfo {author} {\bibfnamefont {T.~T.}\ \bibnamefont
  {{Nakamura}}}\ and\ \bibinfo {author} {\bibfnamefont {S.}~\bibnamefont
  {{Deguchi}}},\ }\href {https://doi.org/10.1143/PTPS.133.137} {\bibfield
  {journal} {\bibinfo  {journal} {Progress of Theoretical Physics Supplement}\
  }\textbf {\bibinfo {volume} {133}},\ \bibinfo {pages} {137} (\bibinfo {year}
  {1999})}\BibitemShut {NoStop}%
\bibitem [{\citenamefont {Nakamura}(1998)}]{PhysRevLett.80.1138}%
  \BibitemOpen
  \bibfield  {author} {\bibinfo {author} {\bibfnamefont {T.~T.}\ \bibnamefont
  {Nakamura}},\ }\href {https://doi.org/10.1103/PhysRevLett.80.1138} {\bibfield
   {journal} {\bibinfo  {journal} {Phys. Rev. Lett.}\ }\textbf {\bibinfo
  {volume} {80}},\ \bibinfo {pages} {1138} (\bibinfo {year}
  {1998})}\BibitemShut {NoStop}%
\bibitem [{\citenamefont {Takahashi}\ and\ \citenamefont
  {Nakamura}(2003)}]{Takahashi_2003}%
  \BibitemOpen
  \bibfield  {author} {\bibinfo {author} {\bibfnamefont {R.}~\bibnamefont
  {Takahashi}}\ and\ \bibinfo {author} {\bibfnamefont {T.}~\bibnamefont
  {Nakamura}},\ }\href {https://doi.org/10.1086/377430} {\bibfield  {journal}
  {\bibinfo  {journal} {The Astrophysical Journal}\ }\textbf {\bibinfo {volume}
  {595}},\ \bibinfo {pages} {1039–1051} (\bibinfo {year} {2003})}\BibitemShut
  {NoStop}%
\bibitem [{\citenamefont {Oguri}(2018)}]{Oguri:2018muv}%
  \BibitemOpen
  \bibfield  {author} {\bibinfo {author} {\bibfnamefont {M.}~\bibnamefont
  {Oguri}},\ }\href {https://doi.org/10.1093/mnras/sty2145} {\bibfield
  {journal} {\bibinfo  {journal} {Mon. Not. Roy. Astron. Soc.}\ }\textbf
  {\bibinfo {volume} {480}},\ \bibinfo {pages} {3842} (\bibinfo {year}
  {2018})},\ \Eprint {https://arxiv.org/abs/1807.02584} {arXiv:1807.02584
  [astro-ph.CO]} \BibitemShut {NoStop}%
\bibitem [{\citenamefont {Li}\ \emph {et~al.}(2018)\citenamefont {Li},
  \citenamefont {Mao}, \citenamefont {Zhao},\ and\ \citenamefont
  {Lu}}]{Li:2018prc}%
  \BibitemOpen
  \bibfield  {author} {\bibinfo {author} {\bibfnamefont {S.-S.}\ \bibnamefont
  {Li}}, \bibinfo {author} {\bibfnamefont {S.}~\bibnamefont {Mao}}, \bibinfo
  {author} {\bibfnamefont {Y.}~\bibnamefont {Zhao}},\ and\ \bibinfo {author}
  {\bibfnamefont {Y.}~\bibnamefont {Lu}},\ }\href
  {https://doi.org/10.1093/mnras/sty411} {\bibfield  {journal} {\bibinfo
  {journal} {Mon. Not. Roy. Astron. Soc.}\ }\textbf {\bibinfo {volume} {476}},\
  \bibinfo {pages} {2220} (\bibinfo {year} {2018})},\ \Eprint
  {https://arxiv.org/abs/1802.05089} {arXiv:1802.05089 [astro-ph.CO]}
  \BibitemShut {NoStop}%
\bibitem [{\citenamefont {Evans}\ \emph {et~al.}(2021)\citenamefont {Evans},
  \citenamefont {Adhikari}, \citenamefont {Afle}, \citenamefont {Ballmer},
  \citenamefont {Biscoveanu}, \citenamefont {Borhanian}, \citenamefont {Brown},
  \citenamefont {Chen}, \citenamefont {Eisenstein}, \citenamefont {Gruson}
  \emph {et~al.}}]{evans2021horizon}%
  \BibitemOpen
  \bibfield  {author} {\bibinfo {author} {\bibfnamefont {M.}~\bibnamefont
  {Evans}}, \bibinfo {author} {\bibfnamefont {R.~X.}\ \bibnamefont {Adhikari}},
  \bibinfo {author} {\bibfnamefont {C.}~\bibnamefont {Afle}}, \bibinfo {author}
  {\bibfnamefont {S.~W.}\ \bibnamefont {Ballmer}}, \bibinfo {author}
  {\bibfnamefont {S.}~\bibnamefont {Biscoveanu}}, \bibinfo {author}
  {\bibfnamefont {S.}~\bibnamefont {Borhanian}}, \bibinfo {author}
  {\bibfnamefont {D.~A.}\ \bibnamefont {Brown}}, \bibinfo {author}
  {\bibfnamefont {Y.}~\bibnamefont {Chen}}, \bibinfo {author} {\bibfnamefont
  {R.}~\bibnamefont {Eisenstein}}, \bibinfo {author} {\bibfnamefont
  {A.}~\bibnamefont {Gruson}}, \emph {et~al.},\ }\href@noop {} {\bibfield
  {journal} {\bibinfo  {journal} {arXiv preprint arXiv:2109.09882}\ } (\bibinfo
  {year} {2021})}\BibitemShut {NoStop}%
\bibitem [{\citenamefont {Maggiore}\ \emph {et~al.}(2020)\citenamefont
  {Maggiore} \emph {et~al.}}]{Maggiore:2019uih}%
  \BibitemOpen
  \bibfield  {author} {\bibinfo {author} {\bibfnamefont {M.}~\bibnamefont
  {Maggiore}} \emph {et~al.},\ }\href
  {https://doi.org/10.1088/1475-7516/2020/03/050} {\bibfield  {journal}
  {\bibinfo  {journal} {JCAP}\ }\textbf {\bibinfo {volume} {03}},\ \bibinfo
  {pages} {050}},\ \Eprint {https://arxiv.org/abs/1912.02622} {arXiv:1912.02622
  [astro-ph.CO]} \BibitemShut {NoStop}%
\bibitem [{\citenamefont {Kawamura}\ \emph {et~al.}(2021)\citenamefont
  {Kawamura} \emph {et~al.}}]{Kawamura:2020pcg}%
  \BibitemOpen
  \bibfield  {author} {\bibinfo {author} {\bibfnamefont {S.}~\bibnamefont
  {Kawamura}} \emph {et~al.},\ }\href {https://doi.org/10.1093/ptep/ptab019}
  {\bibfield  {journal} {\bibinfo  {journal} {PTEP}\ }\textbf {\bibinfo
  {volume} {2021}},\ \bibinfo {pages} {05A105} (\bibinfo {year} {2021})},\
  \Eprint {https://arxiv.org/abs/2006.13545} {arXiv:2006.13545 [gr-qc]}
  \BibitemShut {NoStop}%
\bibitem [{\citenamefont {Barausse}\ \emph {et~al.}(2020)\citenamefont
  {Barausse} \emph {et~al.}}]{Barausse:2020rsu}%
  \BibitemOpen
  \bibfield  {author} {\bibinfo {author} {\bibfnamefont {E.}~\bibnamefont
  {Barausse}} \emph {et~al.},\ }\href
  {https://doi.org/10.1007/s10714-020-02691-1} {\bibfield  {journal} {\bibinfo
  {journal} {Gen. Rel. Grav.}\ }\textbf {\bibinfo {volume} {52}},\ \bibinfo
  {pages} {81} (\bibinfo {year} {2020})},\ \Eprint
  {https://arxiv.org/abs/2001.09793} {arXiv:2001.09793 [gr-qc]} \BibitemShut
  {NoStop}%
\bibitem [{\citenamefont {Lai}\ \emph {et~al.}(2018)\citenamefont {Lai},
  \citenamefont {Hannuksela}, \citenamefont {Herrera-Mart\'\i{}n},
  \citenamefont {Diego}, \citenamefont {Broadhurst},\ and\ \citenamefont
  {Li}}]{Lai:2018rto}%
  \BibitemOpen
  \bibfield  {author} {\bibinfo {author} {\bibfnamefont {K.-H.}\ \bibnamefont
  {Lai}}, \bibinfo {author} {\bibfnamefont {O.~A.}\ \bibnamefont {Hannuksela}},
  \bibinfo {author} {\bibfnamefont {A.}~\bibnamefont {Herrera-Mart\'\i{}n}},
  \bibinfo {author} {\bibfnamefont {J.~M.}\ \bibnamefont {Diego}}, \bibinfo
  {author} {\bibfnamefont {T.}~\bibnamefont {Broadhurst}},\ and\ \bibinfo
  {author} {\bibfnamefont {T.~G.~F.}\ \bibnamefont {Li}},\ }\href
  {https://doi.org/10.1103/PhysRevD.98.083005} {\bibfield  {journal} {\bibinfo
  {journal} {Phys. Rev. D}\ }\textbf {\bibinfo {volume} {98}},\ \bibinfo
  {pages} {083005} (\bibinfo {year} {2018})},\ \Eprint
  {https://arxiv.org/abs/1801.07840} {arXiv:1801.07840 [gr-qc]} \BibitemShut
  {NoStop}%
\bibitem [{\citenamefont {Diego}(2020)}]{Diego:2019rzc}%
  \BibitemOpen
  \bibfield  {author} {\bibinfo {author} {\bibfnamefont {J.~M.}\ \bibnamefont
  {Diego}},\ }\href {https://doi.org/10.1103/PhysRevD.101.123512} {\bibfield
  {journal} {\bibinfo  {journal} {Phys. Rev. D}\ }\textbf {\bibinfo {volume}
  {101}},\ \bibinfo {pages} {123512} (\bibinfo {year} {2020})},\ \Eprint
  {https://arxiv.org/abs/1911.05736} {arXiv:1911.05736 [astro-ph.CO]}
  \BibitemShut {NoStop}%
\bibitem [{\citenamefont {Oguri}\ and\ \citenamefont
  {Takahashi}(2020)}]{Oguri:2020ldf}%
  \BibitemOpen
  \bibfield  {author} {\bibinfo {author} {\bibfnamefont {M.}~\bibnamefont
  {Oguri}}\ and\ \bibinfo {author} {\bibfnamefont {R.}~\bibnamefont
  {Takahashi}},\ }\href {https://doi.org/10.3847/1538-4357/abafab} {\bibfield
  {journal} {\bibinfo  {journal} {Astrophys. J.}\ }\textbf {\bibinfo {volume}
  {901}},\ \bibinfo {pages} {58} (\bibinfo {year} {2020})},\ \Eprint
  {https://arxiv.org/abs/2007.01936} {arXiv:2007.01936 [astro-ph.CO]}
  \BibitemShut {NoStop}%
\bibitem [{\citenamefont {{Xu}}\ \emph {et~al.}(2022)\citenamefont {{Xu}},
  \citenamefont {{Ezquiaga}},\ and\ \citenamefont
  {{Holz}}}]{2022ApJ...929....9X}%
  \BibitemOpen
  \bibfield  {author} {\bibinfo {author} {\bibfnamefont {F.}~\bibnamefont
  {{Xu}}}, \bibinfo {author} {\bibfnamefont {J.~M.}\ \bibnamefont
  {{Ezquiaga}}},\ and\ \bibinfo {author} {\bibfnamefont {D.~E.}\ \bibnamefont
  {{Holz}}},\ }\href {https://doi.org/10.3847/1538-4357/ac58f8} {\bibfield
  {journal} {\bibinfo  {journal} {\apj}\ }\textbf {\bibinfo {volume} {929}},\
  \bibinfo {eid} {9} (\bibinfo {year} {2022})},\ \Eprint
  {https://arxiv.org/abs/2105.14390} {arXiv:2105.14390 [astro-ph.CO]}
  \BibitemShut {NoStop}%
\bibitem [{\citenamefont {Yu}\ \emph {et~al.}(2020)\citenamefont {Yu},
  \citenamefont {Zhang},\ and\ \citenamefont {Wang}}]{Yu:2020agu}%
  \BibitemOpen
  \bibfield  {author} {\bibinfo {author} {\bibfnamefont {H.}~\bibnamefont
  {Yu}}, \bibinfo {author} {\bibfnamefont {P.}~\bibnamefont {Zhang}},\ and\
  \bibinfo {author} {\bibfnamefont {F.-Y.}\ \bibnamefont {Wang}},\ }\href
  {https://doi.org/10.1093/mnras/staa1952} {\bibfield  {journal} {\bibinfo
  {journal} {Mon. Not. Roy. Astron. Soc.}\ }\textbf {\bibinfo {volume} {497}},\
  \bibinfo {pages} {204} (\bibinfo {year} {2020})},\ \Eprint
  {https://arxiv.org/abs/2007.00828} {arXiv:2007.00828 [astro-ph.CO]}
  \BibitemShut {NoStop}%
\bibitem [{\citenamefont {Hannuksela}\ \emph {et~al.}(2020)\citenamefont
  {Hannuksela}, \citenamefont {Collett}, \citenamefont {\c{C}al\i{}\c{s}kan},\
  and\ \citenamefont {Li}}]{Hannuksela:2020xor}%
  \BibitemOpen
  \bibfield  {author} {\bibinfo {author} {\bibfnamefont {O.~A.}\ \bibnamefont
  {Hannuksela}}, \bibinfo {author} {\bibfnamefont {T.~E.}\ \bibnamefont
  {Collett}}, \bibinfo {author} {\bibfnamefont {M.}~\bibnamefont
  {\c{C}al\i{}\c{s}kan}},\ and\ \bibinfo {author} {\bibfnamefont {T.~G.~F.}\
  \bibnamefont {Li}},\ }\href {https://doi.org/10.1093/mnras/staa2577}
  {\bibfield  {journal} {\bibinfo  {journal} {Mon. Not. Roy. Astron. Soc.}\
  }\textbf {\bibinfo {volume} {498}},\ \bibinfo {pages} {3395} (\bibinfo {year}
  {2020})},\ \Eprint {https://arxiv.org/abs/2004.13811} {arXiv:2004.13811
  [astro-ph.HE]} \BibitemShut {NoStop}%
\bibitem [{\citenamefont {Sereno}\ \emph {et~al.}(2011)\citenamefont {Sereno},
  \citenamefont {Jetzer}, \citenamefont {Sesana},\ and\ \citenamefont
  {Volonteri}}]{Sereno:2011ty}%
  \BibitemOpen
  \bibfield  {author} {\bibinfo {author} {\bibfnamefont {M.}~\bibnamefont
  {Sereno}}, \bibinfo {author} {\bibfnamefont {P.}~\bibnamefont {Jetzer}},
  \bibinfo {author} {\bibfnamefont {A.}~\bibnamefont {Sesana}},\ and\ \bibinfo
  {author} {\bibfnamefont {M.}~\bibnamefont {Volonteri}},\ }\href
  {https://doi.org/10.1111/j.1365-2966.2011.18895.x} {\bibfield  {journal}
  {\bibinfo  {journal} {Mon. Not. Roy. Astron. Soc.}\ }\textbf {\bibinfo
  {volume} {415}},\ \bibinfo {pages} {2773} (\bibinfo {year} {2011})},\ \Eprint
  {https://arxiv.org/abs/1104.1977} {arXiv:1104.1977 [astro-ph.CO]}
  \BibitemShut {NoStop}%
\bibitem [{\citenamefont {Liao}\ \emph {et~al.}(2017)\citenamefont {Liao},
  \citenamefont {Fan}, \citenamefont {Ding}, \citenamefont {Biesiada},\ and\
  \citenamefont {Zhu}}]{Liao:2017ioi}%
  \BibitemOpen
  \bibfield  {author} {\bibinfo {author} {\bibfnamefont {K.}~\bibnamefont
  {Liao}}, \bibinfo {author} {\bibfnamefont {X.-L.}\ \bibnamefont {Fan}},
  \bibinfo {author} {\bibfnamefont {X.-H.}\ \bibnamefont {Ding}}, \bibinfo
  {author} {\bibfnamefont {M.}~\bibnamefont {Biesiada}},\ and\ \bibinfo
  {author} {\bibfnamefont {Z.-H.}\ \bibnamefont {Zhu}},\ }\href
  {https://doi.org/10.1038/s41467-017-01152-9} {\bibfield  {journal} {\bibinfo
  {journal} {Nature Commun.}\ }\textbf {\bibinfo {volume} {8}},\ \bibinfo
  {pages} {1148} (\bibinfo {year} {2017})},\ \bibinfo {note} {[Erratum: Nature
  Commun. 8, 2136 (2017)]},\ \Eprint {https://arxiv.org/abs/1703.04151}
  {arXiv:1703.04151 [astro-ph.CO]} \BibitemShut {NoStop}%
\bibitem [{\citenamefont {Cao}\ \emph {et~al.}(2019)\citenamefont {Cao},
  \citenamefont {Qi}, \citenamefont {Cao}, \citenamefont {Biesiada},
  \citenamefont {Li}, \citenamefont {Pan},\ and\ \citenamefont
  {Zhu}}]{Cao:2019kgn}%
  \BibitemOpen
  \bibfield  {author} {\bibinfo {author} {\bibfnamefont {S.}~\bibnamefont
  {Cao}}, \bibinfo {author} {\bibfnamefont {J.}~\bibnamefont {Qi}}, \bibinfo
  {author} {\bibfnamefont {Z.}~\bibnamefont {Cao}}, \bibinfo {author}
  {\bibfnamefont {M.}~\bibnamefont {Biesiada}}, \bibinfo {author}
  {\bibfnamefont {J.}~\bibnamefont {Li}}, \bibinfo {author} {\bibfnamefont
  {Y.}~\bibnamefont {Pan}},\ and\ \bibinfo {author} {\bibfnamefont {Z.-H.}\
  \bibnamefont {Zhu}},\ }\href {https://doi.org/10.1038/s41598-019-47616-4}
  {\bibfield  {journal} {\bibinfo  {journal} {Sci. Rep.}\ }\textbf {\bibinfo
  {volume} {9}},\ \bibinfo {pages} {11608} (\bibinfo {year} {2019})},\ \Eprint
  {https://arxiv.org/abs/1910.10365} {arXiv:1910.10365 [astro-ph.CO]}
  \BibitemShut {NoStop}%
\bibitem [{\citenamefont {Li}\ \emph {et~al.}(2019)\citenamefont {Li},
  \citenamefont {Fan},\ and\ \citenamefont {Gou}}]{Li:2019rns}%
  \BibitemOpen
  \bibfield  {author} {\bibinfo {author} {\bibfnamefont {Y.}~\bibnamefont
  {Li}}, \bibinfo {author} {\bibfnamefont {X.}~\bibnamefont {Fan}},\ and\
  \bibinfo {author} {\bibfnamefont {L.}~\bibnamefont {Gou}},\ }\href
  {https://doi.org/10.3847/1538-4357/ab037e} {\bibfield  {journal} {\bibinfo
  {journal} {Astrophys. J.}\ }\textbf {\bibinfo {volume} {873}},\ \bibinfo
  {pages} {37} (\bibinfo {year} {2019})},\ \Eprint
  {https://arxiv.org/abs/1901.10638} {arXiv:1901.10638 [astro-ph.CO]}
  \BibitemShut {NoStop}%
\bibitem [{\citenamefont {Wempe}\ \emph {et~al.}(2022)\citenamefont {Wempe},
  \citenamefont {Koopmans}, \citenamefont {Wierda}, \citenamefont
  {Hannuksela},\ and\ \citenamefont {Broeck}}]{wempe2022lensing}%
  \BibitemOpen
  \bibfield  {author} {\bibinfo {author} {\bibfnamefont {E.}~\bibnamefont
  {Wempe}}, \bibinfo {author} {\bibfnamefont {L.~V.}\ \bibnamefont {Koopmans}},
  \bibinfo {author} {\bibfnamefont {A.}~\bibnamefont {Wierda}}, \bibinfo
  {author} {\bibfnamefont {O.~A.}\ \bibnamefont {Hannuksela}},\ and\ \bibinfo
  {author} {\bibfnamefont {C.~v.~d.}\ \bibnamefont {Broeck}},\ }\href@noop {}
  {\bibfield  {journal} {\bibinfo  {journal} {arXiv preprint arXiv:2204.08732}\
  } (\bibinfo {year} {2022})}\BibitemShut {NoStop}%
\bibitem [{\citenamefont {Ali}\ \emph {et~al.}(2023)\citenamefont {Ali},
  \citenamefont {Stoikos}, \citenamefont {Meade}, \citenamefont {Kesden},\ and\
  \citenamefont {King}}]{PhysRevD.107.103023}%
  \BibitemOpen
  \bibfield  {author} {\bibinfo {author} {\bibfnamefont {S.}~\bibnamefont
  {Ali}}, \bibinfo {author} {\bibfnamefont {E.}~\bibnamefont {Stoikos}},
  \bibinfo {author} {\bibfnamefont {E.}~\bibnamefont {Meade}}, \bibinfo
  {author} {\bibfnamefont {M.}~\bibnamefont {Kesden}},\ and\ \bibinfo {author}
  {\bibfnamefont {L.}~\bibnamefont {King}},\ }\href
  {https://doi.org/10.1103/PhysRevD.107.103023} {\bibfield  {journal} {\bibinfo
   {journal} {Phys. Rev. D}\ }\textbf {\bibinfo {volume} {107}},\ \bibinfo
  {pages} {103023} (\bibinfo {year} {2023})}\BibitemShut {NoStop}%
\bibitem [{\citenamefont {{Takahashi}}(2004)}]{Takahashi_2004}%
  \BibitemOpen
  \bibfield  {author} {\bibinfo {author} {\bibfnamefont {R.}~\bibnamefont
  {{Takahashi}}},\ }\href {https://doi.org/10.1051/0004-6361:20040212}
  {\bibfield  {journal} {\bibinfo  {journal} {\aap}\ }\textbf {\bibinfo
  {volume} {423}},\ \bibinfo {pages} {787} (\bibinfo {year} {2004})},\ \Eprint
  {https://arxiv.org/abs/astro-ph/0402165} {arXiv:astro-ph/0402165 [astro-ph]}
  \BibitemShut {NoStop}%
\bibitem [{\citenamefont {{Cheung}}\ \emph {et~al.}(2021)\citenamefont
  {{Cheung}}, \citenamefont {{Gais}}, \citenamefont {{Hannuksela}},\ and\
  \citenamefont {{Li}}}]{2021MNRAS.503.3326C}%
  \BibitemOpen
  \bibfield  {author} {\bibinfo {author} {\bibfnamefont {M.~H.~Y.}\
  \bibnamefont {{Cheung}}}, \bibinfo {author} {\bibfnamefont {J.}~\bibnamefont
  {{Gais}}}, \bibinfo {author} {\bibfnamefont {O.~A.}\ \bibnamefont
  {{Hannuksela}}},\ and\ \bibinfo {author} {\bibfnamefont {T.~G.~F.}\
  \bibnamefont {{Li}}},\ }\href {https://doi.org/10.1093/mnras/stab579}
  {\bibfield  {journal} {\bibinfo  {journal} {\mnras}\ }\textbf {\bibinfo
  {volume} {503}},\ \bibinfo {pages} {3326} (\bibinfo {year} {2021})},\ \Eprint
  {https://arxiv.org/abs/2012.07800} {arXiv:2012.07800 [astro-ph.HE]}
  \BibitemShut {NoStop}%
\bibitem [{\citenamefont {Ezquiaga}\ \emph {et~al.}(2021)\citenamefont
  {Ezquiaga}, \citenamefont {Holz}, \citenamefont {Hu}, \citenamefont {Lagos},\
  and\ \citenamefont {Wald}}]{Ezquiaga_2021}%
  \BibitemOpen
  \bibfield  {author} {\bibinfo {author} {\bibfnamefont {J.~M.}\ \bibnamefont
  {Ezquiaga}}, \bibinfo {author} {\bibfnamefont {D.~E.}\ \bibnamefont {Holz}},
  \bibinfo {author} {\bibfnamefont {W.}~\bibnamefont {Hu}}, \bibinfo {author}
  {\bibfnamefont {M.}~\bibnamefont {Lagos}},\ and\ \bibinfo {author}
  {\bibfnamefont {R.~M.}\ \bibnamefont {Wald}},\ }\bibfield  {journal}
  {\bibinfo  {journal} {Physical Review D}\ }\textbf {\bibinfo {volume}
  {103}},\ \href {https://doi.org/10.1103/physrevd.103.064047}
  {10.1103/physrevd.103.064047} (\bibinfo {year} {2021})\BibitemShut {NoStop}%
\bibitem [{\citenamefont {Chan}\ \emph {et~al.}(2025)\citenamefont {Chan},
  \citenamefont {Seo}, \citenamefont {Li}, \citenamefont {Fong},\ and\
  \citenamefont {Ezquiaga}}]{PhysRevD.111.084019}%
  \BibitemOpen
  \bibfield  {author} {\bibinfo {author} {\bibfnamefont {J.~C.~L.}\
  \bibnamefont {Chan}}, \bibinfo {author} {\bibfnamefont {E.}~\bibnamefont
  {Seo}}, \bibinfo {author} {\bibfnamefont {A.~K.~Y.}\ \bibnamefont {Li}},
  \bibinfo {author} {\bibfnamefont {H.}~\bibnamefont {Fong}},\ and\ \bibinfo
  {author} {\bibfnamefont {J.~M.}\ \bibnamefont {Ezquiaga}},\ }\href
  {https://doi.org/10.1103/PhysRevD.111.084019} {\bibfield  {journal} {\bibinfo
   {journal} {Phys. Rev. D}\ }\textbf {\bibinfo {volume} {111}},\ \bibinfo
  {pages} {084019} (\bibinfo {year} {2025})}\BibitemShut {NoStop}%
\bibitem [{\citenamefont {{Ohanian}}(1983)}]{1983ApJ...271..551O}%
  \BibitemOpen
  \bibfield  {author} {\bibinfo {author} {\bibfnamefont {H.~C.}\ \bibnamefont
  {{Ohanian}}},\ }\href {https://doi.org/10.1086/161221} {\bibfield  {journal}
  {\bibinfo  {journal} {\apj}\ }\textbf {\bibinfo {volume} {271}},\ \bibinfo
  {pages} {551} (\bibinfo {year} {1983})}\BibitemShut {NoStop}%
\bibitem [{\citenamefont {{Turner}}\ \emph {et~al.}(1984)\citenamefont
  {{Turner}}, \citenamefont {{Ostriker}},\ and\ \citenamefont
  {{Gott}}}]{1984ApJ...284....1T}%
  \BibitemOpen
  \bibfield  {author} {\bibinfo {author} {\bibfnamefont {E.~L.}\ \bibnamefont
  {{Turner}}}, \bibinfo {author} {\bibfnamefont {J.~P.}\ \bibnamefont
  {{Ostriker}}},\ and\ \bibinfo {author} {\bibfnamefont {I.}~\bibnamefont
  {{Gott}}, \bibfnamefont {J.~R.}},\ }\href {https://doi.org/10.1086/162379}
  {\bibfield  {journal} {\bibinfo  {journal} {\apj}\ }\textbf {\bibinfo
  {volume} {284}},\ \bibinfo {pages} {1} (\bibinfo {year} {1984})}\BibitemShut
  {NoStop}%
\bibitem [{\citenamefont {{Blandford}}\ and\ \citenamefont
  {{Narayan}}(1986)}]{1986ApJ...310..568B}%
  \BibitemOpen
  \bibfield  {author} {\bibinfo {author} {\bibfnamefont {R.}~\bibnamefont
  {{Blandford}}}\ and\ \bibinfo {author} {\bibfnamefont {R.}~\bibnamefont
  {{Narayan}}},\ }\href {https://doi.org/10.1086/164709} {\bibfield  {journal}
  {\bibinfo  {journal} {\apj}\ }\textbf {\bibinfo {volume} {310}},\ \bibinfo
  {pages} {568} (\bibinfo {year} {1986})}\BibitemShut {NoStop}%
\bibitem [{\citenamefont {{Kormann}}\ \emph {et~al.}(1994)\citenamefont
  {{Kormann}}, \citenamefont {{Schneider}},\ and\ \citenamefont
  {{Bartelmann}}}]{1994A&A...284..285K}%
  \BibitemOpen
  \bibfield  {author} {\bibinfo {author} {\bibfnamefont {R.}~\bibnamefont
  {{Kormann}}}, \bibinfo {author} {\bibfnamefont {P.}~\bibnamefont
  {{Schneider}}},\ and\ \bibinfo {author} {\bibfnamefont {M.}~\bibnamefont
  {{Bartelmann}}},\ }\href@noop {} {\bibfield  {journal} {\bibinfo  {journal}
  {\aap}\ }\textbf {\bibinfo {volume} {284}},\ \bibinfo {pages} {285} (\bibinfo
  {year} {1994})}\BibitemShut {NoStop}%
\bibitem [{\citenamefont {{Asada}}\ \emph {et~al.}(2003)\citenamefont
  {{Asada}}, \citenamefont {{Hamana}},\ and\ \citenamefont
  {{Kasai}}}]{2003A&A...397..825A}%
  \BibitemOpen
  \bibfield  {author} {\bibinfo {author} {\bibfnamefont {H.}~\bibnamefont
  {{Asada}}}, \bibinfo {author} {\bibfnamefont {T.}~\bibnamefont {{Hamana}}},\
  and\ \bibinfo {author} {\bibfnamefont {M.}~\bibnamefont {{Kasai}}},\ }\href
  {https://doi.org/10.1051/0004-6361:20021599} {\bibfield  {journal} {\bibinfo
  {journal} {\aap}\ }\textbf {\bibinfo {volume} {397}},\ \bibinfo {pages} {825}
  (\bibinfo {year} {2003})},\ \Eprint {https://arxiv.org/abs/astro-ph/0211389}
  {arXiv:astro-ph/0211389 [astro-ph]} \BibitemShut {NoStop}%
\bibitem [{\citenamefont {{Oguri}}(2010)}]{2010PASJ...62.1017O}%
  \BibitemOpen
  \bibfield  {author} {\bibinfo {author} {\bibfnamefont {M.}~\bibnamefont
  {{Oguri}}},\ }\href {https://doi.org/10.1093/pasj/62.4.1017} {\bibfield
  {journal} {\bibinfo  {journal} {\pasj}\ }\textbf {\bibinfo {volume} {62}},\
  \bibinfo {pages} {1017} (\bibinfo {year} {2010})},\ \Eprint
  {https://arxiv.org/abs/1005.3103} {arXiv:1005.3103 [astro-ph.CO]}
  \BibitemShut {NoStop}%
\bibitem [{\citenamefont {{Browne}}\ \emph {et~al.}(2003)\citenamefont
  {{Browne}}, \citenamefont {{Wilkinson}}, \citenamefont {{Jackson}},
  \citenamefont {{Myers}}, \citenamefont {{Fassnacht}}, \citenamefont
  {{Koopmans}}, \citenamefont {{Marlow}}, \citenamefont {{Norbury}},
  \citenamefont {{Rusin}}, \citenamefont {{Sykes}}, \citenamefont {{Biggs}},
  \citenamefont {{Blandford}}, \citenamefont {{de Bruyn}}, \citenamefont
  {{Chae}}, \citenamefont {{Helbig}}, \citenamefont {{King}}, \citenamefont
  {{McKean}}, \citenamefont {{Pearson}}, \citenamefont {{Phillips}},
  \citenamefont {{Readhead}}, \citenamefont {{Xanthopoulos}},\ and\
  \citenamefont {{York}}}]{2003MNRAS.341...13B}%
  \BibitemOpen
  \bibfield  {author} {\bibinfo {author} {\bibfnamefont {I.~W.~A.}\
  \bibnamefont {{Browne}}}, \bibinfo {author} {\bibfnamefont {P.~N.}\
  \bibnamefont {{Wilkinson}}}, \bibinfo {author} {\bibfnamefont {N.~J.~F.}\
  \bibnamefont {{Jackson}}}, \bibinfo {author} {\bibfnamefont {S.~T.}\
  \bibnamefont {{Myers}}}, \bibinfo {author} {\bibfnamefont {C.~D.}\
  \bibnamefont {{Fassnacht}}}, \bibinfo {author} {\bibfnamefont {L.~V.~E.}\
  \bibnamefont {{Koopmans}}}, \bibinfo {author} {\bibfnamefont {D.~R.}\
  \bibnamefont {{Marlow}}}, \bibinfo {author} {\bibfnamefont {M.}~\bibnamefont
  {{Norbury}}}, \bibinfo {author} {\bibfnamefont {D.}~\bibnamefont {{Rusin}}},
  \bibinfo {author} {\bibfnamefont {C.~M.}\ \bibnamefont {{Sykes}}}, \bibinfo
  {author} {\bibfnamefont {A.~D.}\ \bibnamefont {{Biggs}}}, \bibinfo {author}
  {\bibfnamefont {R.~D.}\ \bibnamefont {{Blandford}}}, \bibinfo {author}
  {\bibfnamefont {A.~G.}\ \bibnamefont {{de Bruyn}}}, \bibinfo {author}
  {\bibfnamefont {K.~H.}\ \bibnamefont {{Chae}}}, \bibinfo {author}
  {\bibfnamefont {P.}~\bibnamefont {{Helbig}}}, \bibinfo {author}
  {\bibfnamefont {L.~J.}\ \bibnamefont {{King}}}, \bibinfo {author}
  {\bibfnamefont {J.~P.}\ \bibnamefont {{McKean}}}, \bibinfo {author}
  {\bibfnamefont {T.~J.}\ \bibnamefont {{Pearson}}}, \bibinfo {author}
  {\bibfnamefont {P.~M.}\ \bibnamefont {{Phillips}}}, \bibinfo {author}
  {\bibfnamefont {A.~C.~S.}\ \bibnamefont {{Readhead}}}, \bibinfo {author}
  {\bibfnamefont {E.}~\bibnamefont {{Xanthopoulos}}},\ and\ \bibinfo {author}
  {\bibfnamefont {T.}~\bibnamefont {{York}}},\ }\href
  {https://doi.org/10.1046/j.1365-8711.2003.06257.x} {\bibfield  {journal}
  {\bibinfo  {journal} {\mnras}\ }\textbf {\bibinfo {volume} {341}},\ \bibinfo
  {pages} {13} (\bibinfo {year} {2003})},\ \Eprint
  {https://arxiv.org/abs/astro-ph/0211069} {arXiv:astro-ph/0211069 [astro-ph]}
  \BibitemShut {NoStop}%
\bibitem [{\citenamefont {{Oguri}}\ \emph {et~al.}(2012)\citenamefont
  {{Oguri}}, \citenamefont {{Inada}}, \citenamefont {{Strauss}}, \citenamefont
  {{Kochanek}}, \citenamefont {{Kayo}}, \citenamefont {{Shin}}, \citenamefont
  {{Morokuma}}, \citenamefont {{Richards}}, \citenamefont {{Rusu}},
  \citenamefont {{Frieman}}, \citenamefont {{Fukugita}}, \citenamefont
  {{Schneider}}, \citenamefont {{York}}, \citenamefont {{Bahcall}},\ and\
  \citenamefont {{White}}}]{2012AJ....143..120O}%
  \BibitemOpen
  \bibfield  {author} {\bibinfo {author} {\bibfnamefont {M.}~\bibnamefont
  {{Oguri}}}, \bibinfo {author} {\bibfnamefont {N.}~\bibnamefont {{Inada}}},
  \bibinfo {author} {\bibfnamefont {M.~A.}\ \bibnamefont {{Strauss}}}, \bibinfo
  {author} {\bibfnamefont {C.~S.}\ \bibnamefont {{Kochanek}}}, \bibinfo
  {author} {\bibfnamefont {I.}~\bibnamefont {{Kayo}}}, \bibinfo {author}
  {\bibfnamefont {M.-S.}\ \bibnamefont {{Shin}}}, \bibinfo {author}
  {\bibfnamefont {T.}~\bibnamefont {{Morokuma}}}, \bibinfo {author}
  {\bibfnamefont {G.~T.}\ \bibnamefont {{Richards}}}, \bibinfo {author}
  {\bibfnamefont {C.~E.}\ \bibnamefont {{Rusu}}}, \bibinfo {author}
  {\bibfnamefont {J.~A.}\ \bibnamefont {{Frieman}}}, \bibinfo {author}
  {\bibfnamefont {M.}~\bibnamefont {{Fukugita}}}, \bibinfo {author}
  {\bibfnamefont {D.~P.}\ \bibnamefont {{Schneider}}}, \bibinfo {author}
  {\bibfnamefont {D.~G.}\ \bibnamefont {{York}}}, \bibinfo {author}
  {\bibfnamefont {N.~A.}\ \bibnamefont {{Bahcall}}},\ and\ \bibinfo {author}
  {\bibfnamefont {R.~L.}\ \bibnamefont {{White}}},\ }\href
  {https://doi.org/10.1088/0004-6256/143/5/120} {\bibfield  {journal} {\bibinfo
   {journal} {\aj}\ }\textbf {\bibinfo {volume} {143}},\ \bibinfo {eid} {120}
  (\bibinfo {year} {2012})},\ \Eprint {https://arxiv.org/abs/1203.1088}
  {arXiv:1203.1088 [astro-ph.CO]} \BibitemShut {NoStop}%
\bibitem [{\citenamefont {{Dux}}\ \emph {et~al.}(2023)\citenamefont {{Dux}},
  \citenamefont {{Lemon}}, \citenamefont {{Courbin}}, \citenamefont {{Sluse}},
  \citenamefont {{Smette}}, \citenamefont {{Anguita}},\ and\ \citenamefont
  {{Neira}}}]{2023arXiv231004494D}%
  \BibitemOpen
  \bibfield  {author} {\bibinfo {author} {\bibfnamefont {F.}~\bibnamefont
  {{Dux}}}, \bibinfo {author} {\bibfnamefont {C.}~\bibnamefont {{Lemon}}},
  \bibinfo {author} {\bibfnamefont {F.}~\bibnamefont {{Courbin}}}, \bibinfo
  {author} {\bibfnamefont {D.}~\bibnamefont {{Sluse}}}, \bibinfo {author}
  {\bibfnamefont {A.}~\bibnamefont {{Smette}}}, \bibinfo {author}
  {\bibfnamefont {T.}~\bibnamefont {{Anguita}}},\ and\ \bibinfo {author}
  {\bibfnamefont {F.}~\bibnamefont {{Neira}}},\ }\href
  {https://doi.org/10.48550/arXiv.2310.04494} {\bibfield  {journal} {\bibinfo
  {journal} {arXiv e-prints}\ ,\ \bibinfo {eid} {arXiv:2310.04494}} (\bibinfo
  {year} {2023})},\ \Eprint {https://arxiv.org/abs/2310.04494}
  {arXiv:2310.04494 [astro-ph.CO]} \BibitemShut {NoStop}%
\bibitem [{\citenamefont {{Richards}}\ \emph {et~al.}(2004)\citenamefont
  {{Richards}}, \citenamefont {{Keeton}}, \citenamefont {{Pindor}},
  \citenamefont {{Hennawi}}, \citenamefont {{Hall}}, \citenamefont {{Turner}},
  \citenamefont {{Inada}}, \citenamefont {{Oguri}}, \citenamefont {{Ichikawa}},
  \citenamefont {{Becker}}, \citenamefont {{Gregg}}, \citenamefont {{White}},
  \citenamefont {{Wyithe}}, \citenamefont {{Schneider}}, \citenamefont
  {{Johnston}}, \citenamefont {{Frieman}},\ and\ \citenamefont
  {{Brinkmann}}}]{2004ApJ...610..679R}%
  \BibitemOpen
  \bibfield  {author} {\bibinfo {author} {\bibfnamefont {G.~T.}\ \bibnamefont
  {{Richards}}}, \bibinfo {author} {\bibfnamefont {C.~R.}\ \bibnamefont
  {{Keeton}}}, \bibinfo {author} {\bibfnamefont {B.}~\bibnamefont {{Pindor}}},
  \bibinfo {author} {\bibfnamefont {J.~F.}\ \bibnamefont {{Hennawi}}}, \bibinfo
  {author} {\bibfnamefont {P.~B.}\ \bibnamefont {{Hall}}}, \bibinfo {author}
  {\bibfnamefont {E.~L.}\ \bibnamefont {{Turner}}}, \bibinfo {author}
  {\bibfnamefont {N.}~\bibnamefont {{Inada}}}, \bibinfo {author} {\bibfnamefont
  {M.}~\bibnamefont {{Oguri}}}, \bibinfo {author} {\bibfnamefont {S.-I.}\
  \bibnamefont {{Ichikawa}}}, \bibinfo {author} {\bibfnamefont {R.~H.}\
  \bibnamefont {{Becker}}}, \bibinfo {author} {\bibfnamefont {M.~D.}\
  \bibnamefont {{Gregg}}}, \bibinfo {author} {\bibfnamefont {R.~L.}\
  \bibnamefont {{White}}}, \bibinfo {author} {\bibfnamefont {J.~S.~B.}\
  \bibnamefont {{Wyithe}}}, \bibinfo {author} {\bibfnamefont {D.~P.}\
  \bibnamefont {{Schneider}}}, \bibinfo {author} {\bibfnamefont {D.~E.}\
  \bibnamefont {{Johnston}}}, \bibinfo {author} {\bibfnamefont {J.~A.}\
  \bibnamefont {{Frieman}}},\ and\ \bibinfo {author} {\bibfnamefont
  {J.}~\bibnamefont {{Brinkmann}}},\ }\href {https://doi.org/10.1086/421868}
  {\bibfield  {journal} {\bibinfo  {journal} {\apj}\ }\textbf {\bibinfo
  {volume} {610}},\ \bibinfo {pages} {679} (\bibinfo {year} {2004})},\ \Eprint
  {https://arxiv.org/abs/astro-ph/0402345} {arXiv:astro-ph/0402345 [astro-ph]}
  \BibitemShut {NoStop}%
\bibitem [{\citenamefont {{Mishra}}\ \emph {et~al.}(2021)\citenamefont
  {{Mishra}}, \citenamefont {{Meena}}, \citenamefont {{More}}, \citenamefont
  {{Bose}},\ and\ \citenamefont {{Bagla}}}]{2021MNRAS.508.4869M}%
  \BibitemOpen
  \bibfield  {author} {\bibinfo {author} {\bibfnamefont {A.}~\bibnamefont
  {{Mishra}}}, \bibinfo {author} {\bibfnamefont {A.~K.}\ \bibnamefont
  {{Meena}}}, \bibinfo {author} {\bibfnamefont {A.}~\bibnamefont {{More}}},
  \bibinfo {author} {\bibfnamefont {S.}~\bibnamefont {{Bose}}},\ and\ \bibinfo
  {author} {\bibfnamefont {J.~S.}\ \bibnamefont {{Bagla}}},\ }\href
  {https://doi.org/10.1093/mnras/stab2875} {\bibfield  {journal} {\bibinfo
  {journal} {\mnras}\ }\textbf {\bibinfo {volume} {508}},\ \bibinfo {pages}
  {4869} (\bibinfo {year} {2021})},\ \Eprint {https://arxiv.org/abs/2102.03946}
  {arXiv:2102.03946 [astro-ph.CO]} \BibitemShut {NoStop}%
\bibitem [{\citenamefont {{More}}\ and\ \citenamefont
  {{More}}(2022)}]{more2022improved}%
  \BibitemOpen
  \bibfield  {author} {\bibinfo {author} {\bibfnamefont {A.}~\bibnamefont
  {{More}}}\ and\ \bibinfo {author} {\bibfnamefont {S.}~\bibnamefont
  {{More}}},\ }\href {https://doi.org/10.1093/mnras/stac1704} {\bibfield
  {journal} {\bibinfo  {journal} {\mnras}\ }\textbf {\bibinfo {volume} {515}},\
  \bibinfo {pages} {1044} (\bibinfo {year} {2022})},\ \Eprint
  {https://arxiv.org/abs/2111.03091} {arXiv:2111.03091 [astro-ph.CO]}
  \BibitemShut {NoStop}%
\bibitem [{\citenamefont {{Haris}}\ \emph {et~al.}(2018)\citenamefont
  {{Haris}}, \citenamefont {{Mehta}}, \citenamefont {{Kumar}}, \citenamefont
  {{Venumadhav}},\ and\ \citenamefont {{Ajith}}}]{2018arXiv180707062H}%
  \BibitemOpen
  \bibfield  {author} {\bibinfo {author} {\bibfnamefont {K.}~\bibnamefont
  {{Haris}}}, \bibinfo {author} {\bibfnamefont {A.~K.}\ \bibnamefont
  {{Mehta}}}, \bibinfo {author} {\bibfnamefont {S.}~\bibnamefont {{Kumar}}},
  \bibinfo {author} {\bibfnamefont {T.}~\bibnamefont {{Venumadhav}}},\ and\
  \bibinfo {author} {\bibfnamefont {P.}~\bibnamefont {{Ajith}}},\ }\href
  {https://doi.org/10.48550/arXiv.1807.07062} {\bibfield  {journal} {\bibinfo
  {journal} {arXiv e-prints}\ ,\ \bibinfo {eid} {arXiv:1807.07062}} (\bibinfo
  {year} {2018})},\ \Eprint {https://arxiv.org/abs/1807.07062}
  {arXiv:1807.07062 [gr-qc]} \BibitemShut {NoStop}%
\bibitem [{\citenamefont {{Guo}}\ and\ \citenamefont
  {{Lu}}(2020)}]{2020PhRvD.102l4076G}%
  \BibitemOpen
  \bibfield  {author} {\bibinfo {author} {\bibfnamefont {X.}~\bibnamefont
  {{Guo}}}\ and\ \bibinfo {author} {\bibfnamefont {Y.}~\bibnamefont {{Lu}}},\
  }\href {https://doi.org/10.1103/PhysRevD.102.124076} {\bibfield  {journal}
  {\bibinfo  {journal} {\prd}\ }\textbf {\bibinfo {volume} {102}},\ \bibinfo
  {eid} {124076} (\bibinfo {year} {2020})},\ \Eprint
  {https://arxiv.org/abs/2012.03474} {arXiv:2012.03474 [gr-qc]} \BibitemShut
  {NoStop}%
\bibitem [{\citenamefont {{Moylan}}\ \emph {et~al.}(2008)\citenamefont
  {{Moylan}}, \citenamefont {{McClelland}}, \citenamefont {{Scott}},
  \citenamefont {{Searle}},\ and\ \citenamefont
  {{Bicknell}}}]{2008mgm..conf..807M}%
  \BibitemOpen
  \bibfield  {author} {\bibinfo {author} {\bibfnamefont {A.~J.}\ \bibnamefont
  {{Moylan}}}, \bibinfo {author} {\bibfnamefont {D.~E.}\ \bibnamefont
  {{McClelland}}}, \bibinfo {author} {\bibfnamefont {S.~M.}\ \bibnamefont
  {{Scott}}}, \bibinfo {author} {\bibfnamefont {A.~C.}\ \bibnamefont
  {{Searle}}},\ and\ \bibinfo {author} {\bibfnamefont {G.~V.}\ \bibnamefont
  {{Bicknell}}},\ }in\ \href {https://doi.org/10.1142/9789812834300_0038}
  {\emph {\bibinfo {booktitle} {The Eleventh Marcel Grossmann Meeting On Recent
  Developments in Theoretical and Experimental General Relativity, Gravitation
  and Relativistic Field Theories}}}\ (\bibinfo {year} {2008})\ pp.\ \bibinfo
  {pages} {807--823},\ \Eprint {https://arxiv.org/abs/0710.3140}
  {arXiv:0710.3140 [gr-qc]} \BibitemShut {NoStop}%
\bibitem [{\citenamefont {{Grillo}}\ and\ \citenamefont
  {{Cordes}}(2018)}]{2018arXiv181009058G}%
  \BibitemOpen
  \bibfield  {author} {\bibinfo {author} {\bibfnamefont {G.}~\bibnamefont
  {{Grillo}}}\ and\ \bibinfo {author} {\bibfnamefont {J.}~\bibnamefont
  {{Cordes}}},\ }\href {https://doi.org/10.48550/arXiv.1810.09058} {\bibfield
  {journal} {\bibinfo  {journal} {arXiv e-prints}\ ,\ \bibinfo {eid}
  {arXiv:1810.09058}} (\bibinfo {year} {2018})},\ \Eprint
  {https://arxiv.org/abs/1810.09058} {arXiv:1810.09058 [astro-ph.CO]}
  \BibitemShut {NoStop}%
\bibitem [{\citenamefont {{Feldbrugge}}\ \emph {et~al.}(2019)\citenamefont
  {{Feldbrugge}}, \citenamefont {{Pen}},\ and\ \citenamefont
  {{Turok}}}]{2019arXiv190904632F}%
  \BibitemOpen
  \bibfield  {author} {\bibinfo {author} {\bibfnamefont {J.}~\bibnamefont
  {{Feldbrugge}}}, \bibinfo {author} {\bibfnamefont {U.-L.}\ \bibnamefont
  {{Pen}}},\ and\ \bibinfo {author} {\bibfnamefont {N.}~\bibnamefont
  {{Turok}}},\ }\href {https://doi.org/10.48550/arXiv.1909.04632} {\bibfield
  {journal} {\bibinfo  {journal} {arXiv e-prints}\ ,\ \bibinfo {eid}
  {arXiv:1909.04632}} (\bibinfo {year} {2019})},\ \Eprint
  {https://arxiv.org/abs/1909.04632} {arXiv:1909.04632 [astro-ph.HE]}
  \BibitemShut {NoStop}%
\bibitem [{\citenamefont {{Jow}}\ \emph {et~al.}(2023)\citenamefont {{Jow}},
  \citenamefont {{Pen}},\ and\ \citenamefont
  {{Feldbrugge}}}]{2023MNRAS.525.2107J}%
  \BibitemOpen
  \bibfield  {author} {\bibinfo {author} {\bibfnamefont {D.~L.}\ \bibnamefont
  {{Jow}}}, \bibinfo {author} {\bibfnamefont {U.-L.}\ \bibnamefont {{Pen}}},\
  and\ \bibinfo {author} {\bibfnamefont {J.}~\bibnamefont {{Feldbrugge}}},\
  }\href {https://doi.org/10.1093/mnras/stad2332} {\bibfield  {journal}
  {\bibinfo  {journal} {\mnras}\ }\textbf {\bibinfo {volume} {525}},\ \bibinfo
  {pages} {2107} (\bibinfo {year} {2023})},\ \Eprint
  {https://arxiv.org/abs/2204.12004} {arXiv:2204.12004 [astro-ph.HE]}
  \BibitemShut {NoStop}%
\bibitem [{\citenamefont {Takahashi}(2004)}]{Takahashi:2004mc}%
  \BibitemOpen
  \bibfield  {author} {\bibinfo {author} {\bibfnamefont {R.}~\bibnamefont
  {Takahashi}},\ }\href {https://doi.org/10.1051/0004-6361:20040212} {\bibfield
   {journal} {\bibinfo  {journal} {Astron. Astrophys.}\ }\textbf {\bibinfo
  {volume} {423}},\ \bibinfo {pages} {787} (\bibinfo {year} {2004})},\ \Eprint
  {https://arxiv.org/abs/astro-ph/0402165} {arXiv:astro-ph/0402165}
  \BibitemShut {NoStop}%
\bibitem [{\citenamefont {{Takahashi}}(2004)}]{2004A&A...423..787T}%
  \BibitemOpen
  \bibfield  {author} {\bibinfo {author} {\bibfnamefont {R.}~\bibnamefont
  {{Takahashi}}},\ }\href {https://doi.org/10.1051/0004-6361:20040212}
  {\bibfield  {journal} {\bibinfo  {journal} {\aap}\ }\textbf {\bibinfo
  {volume} {423}},\ \bibinfo {pages} {787} (\bibinfo {year} {2004})},\ \Eprint
  {https://arxiv.org/abs/astro-ph/0402165} {arXiv:astro-ph/0402165 [astro-ph]}
  \BibitemShut {NoStop}%
\bibitem [{\citenamefont {{Schneider}}\ \emph {et~al.}(1992)\citenamefont
  {{Schneider}}, \citenamefont {{Ehlers}},\ and\ \citenamefont
  {{Falco}}}]{1992grle.book.....S}%
  \BibitemOpen
  \bibfield  {author} {\bibinfo {author} {\bibfnamefont {P.}~\bibnamefont
  {{Schneider}}}, \bibinfo {author} {\bibfnamefont {J.}~\bibnamefont
  {{Ehlers}}},\ and\ \bibinfo {author} {\bibfnamefont {E.~E.}\ \bibnamefont
  {{Falco}}},\ }\href {https://doi.org/10.1007/978-3-662-03758-4} {\emph
  {\bibinfo {title} {{Gravitational Lenses}}}}\ (\bibinfo {year}
  {1992})\BibitemShut {NoStop}%
\bibitem [{\citenamefont {Matsunaga}\ and\ \citenamefont
  {Yamamoto}(2006)}]{Matsunaga_2006}%
  \BibitemOpen
  \bibfield  {author} {\bibinfo {author} {\bibfnamefont {N.}~\bibnamefont
  {Matsunaga}}\ and\ \bibinfo {author} {\bibfnamefont {K.}~\bibnamefont
  {Yamamoto}},\ }\href {https://doi.org/10.1088/1475-7516/2006/01/023}
  {\bibfield  {journal} {\bibinfo  {journal} {Journal of Cosmology and
  Astroparticle Physics}\ }\textbf {\bibinfo {volume} {2006}}\bibinfo  {number}
  { (01)},\ \bibinfo {pages} {023}}\BibitemShut {NoStop}%
\bibitem [{\citenamefont {Cutler}\ and\ \citenamefont
  {Flanagan}(1994)}]{PhysRevD.49.2658}%
  \BibitemOpen
\bibfield  {number} {  }\bibfield  {author} {\bibinfo {author} {\bibfnamefont
  {C.}~\bibnamefont {Cutler}}\ and\ \bibinfo {author} {\bibfnamefont {E.~E.}\
  \bibnamefont {Flanagan}},\ }\href {https://doi.org/10.1103/PhysRevD.49.2658}
  {\bibfield  {journal} {\bibinfo  {journal} {Phys. Rev. D}\ }\textbf {\bibinfo
  {volume} {49}},\ \bibinfo {pages} {2658} (\bibinfo {year}
  {1994})}\BibitemShut {NoStop}%
\bibitem [{\citenamefont {{M{\"o}ller}}\ \emph {et~al.}(2007)\citenamefont
  {{M{\"o}ller}}, \citenamefont {{Kitzbichler}},\ and\ \citenamefont
  {{Natarajan}}}]{2007MNRAS.379.1195M}%
  \BibitemOpen
  \bibfield  {author} {\bibinfo {author} {\bibfnamefont {O.}~\bibnamefont
  {{M{\"o}ller}}}, \bibinfo {author} {\bibfnamefont {M.}~\bibnamefont
  {{Kitzbichler}}},\ and\ \bibinfo {author} {\bibfnamefont {P.}~\bibnamefont
  {{Natarajan}}},\ }\href {https://doi.org/10.1111/j.1365-2966.2007.12004.x}
  {\bibfield  {journal} {\bibinfo  {journal} {\mnras}\ }\textbf {\bibinfo
  {volume} {379}},\ \bibinfo {pages} {1195} (\bibinfo {year} {2007})},\ \Eprint
  {https://arxiv.org/abs/astro-ph/0607032} {arXiv:astro-ph/0607032 [astro-ph]}
  \BibitemShut {NoStop}%
\bibitem [{\citenamefont {{Keeton}}\ \emph {et~al.}(2000)\citenamefont
  {{Keeton}}, \citenamefont {{Mao}},\ and\ \citenamefont
  {{Witt}}}]{2000ApJ...537..697K}%
  \BibitemOpen
  \bibfield  {author} {\bibinfo {author} {\bibfnamefont {C.~R.}\ \bibnamefont
  {{Keeton}}}, \bibinfo {author} {\bibfnamefont {S.}~\bibnamefont {{Mao}}},\
  and\ \bibinfo {author} {\bibfnamefont {H.~J.}\ \bibnamefont {{Witt}}},\
  }\href {https://doi.org/10.1086/309087} {\bibfield  {journal} {\bibinfo
  {journal} {\apj}\ }\textbf {\bibinfo {volume} {537}},\ \bibinfo {pages} {697}
  (\bibinfo {year} {2000})},\ \Eprint {https://arxiv.org/abs/astro-ph/0002401}
  {arXiv:astro-ph/0002401 [astro-ph]} \BibitemShut {NoStop}%
\bibitem [{\citenamefont {Nitz}\ \emph {et~al.}(2022)\citenamefont {Nitz},
  \citenamefont {Harry}, \citenamefont {Brown}, \citenamefont {Biwer},
  \citenamefont {Willis}, \citenamefont {Canton}, \citenamefont {Capano},
  \citenamefont {Dent}, \citenamefont {Pekowsky}, \citenamefont {Williamson},
  \citenamefont {De}, \citenamefont {Cabero}, \citenamefont {Machenschalk},
  \citenamefont {Macleod}, \citenamefont {Kumar}, \citenamefont {Reyes},
  \citenamefont {dfinstad}, \citenamefont {Pannarale}, \citenamefont {Kumar},
  \citenamefont {Massinger}, \citenamefont {Tápai}, \citenamefont {Singer},
  \citenamefont {Davies}, \citenamefont {Khan}, \citenamefont {Fairhurst},
  \citenamefont {Nielsen}, \citenamefont {Singh}, \citenamefont {Chandra},
  \citenamefont {shasvath},\ and\ \citenamefont {veronica
  villa}}]{alex_nitz_2022_6324278}%
  \BibitemOpen
  \bibfield  {author} {\bibinfo {author} {\bibfnamefont {A.}~\bibnamefont
  {Nitz}}, \bibinfo {author} {\bibfnamefont {I.}~\bibnamefont {Harry}},
  \bibinfo {author} {\bibfnamefont {D.}~\bibnamefont {Brown}}, \bibinfo
  {author} {\bibfnamefont {C.~M.}\ \bibnamefont {Biwer}}, \bibinfo {author}
  {\bibfnamefont {J.}~\bibnamefont {Willis}}, \bibinfo {author} {\bibfnamefont
  {T.~D.}\ \bibnamefont {Canton}}, \bibinfo {author} {\bibfnamefont
  {C.}~\bibnamefont {Capano}}, \bibinfo {author} {\bibfnamefont
  {T.}~\bibnamefont {Dent}}, \bibinfo {author} {\bibfnamefont {L.}~\bibnamefont
  {Pekowsky}}, \bibinfo {author} {\bibfnamefont {A.~R.}\ \bibnamefont
  {Williamson}}, \bibinfo {author} {\bibfnamefont {S.}~\bibnamefont {De}},
  \bibinfo {author} {\bibfnamefont {M.}~\bibnamefont {Cabero}}, \bibinfo
  {author} {\bibfnamefont {B.}~\bibnamefont {Machenschalk}}, \bibinfo {author}
  {\bibfnamefont {D.}~\bibnamefont {Macleod}}, \bibinfo {author} {\bibfnamefont
  {P.}~\bibnamefont {Kumar}}, \bibinfo {author} {\bibfnamefont
  {S.}~\bibnamefont {Reyes}}, \bibinfo {author} {\bibnamefont {dfinstad}},
  \bibinfo {author} {\bibfnamefont {F.}~\bibnamefont {Pannarale}}, \bibinfo
  {author} {\bibfnamefont {S.}~\bibnamefont {Kumar}}, \bibinfo {author}
  {\bibfnamefont {T.}~\bibnamefont {Massinger}}, \bibinfo {author}
  {\bibfnamefont {M.}~\bibnamefont {Tápai}}, \bibinfo {author} {\bibfnamefont
  {L.}~\bibnamefont {Singer}}, \bibinfo {author} {\bibfnamefont {G.~S.~C.}\
  \bibnamefont {Davies}}, \bibinfo {author} {\bibfnamefont {S.}~\bibnamefont
  {Khan}}, \bibinfo {author} {\bibfnamefont {S.}~\bibnamefont {Fairhurst}},
  \bibinfo {author} {\bibfnamefont {A.}~\bibnamefont {Nielsen}}, \bibinfo
  {author} {\bibfnamefont {S.}~\bibnamefont {Singh}}, \bibinfo {author}
  {\bibfnamefont {K.}~\bibnamefont {Chandra}}, \bibinfo {author} {\bibnamefont
  {shasvath}},\ and\ \bibinfo {author} {\bibnamefont {veronica villa}},\ }\href
  {https://doi.org/10.5281/zenodo.6324278} {\bibinfo {title} {gwastro/pycbc:
  v2.0.2 release of pycbc}} (\bibinfo {year} {2022})\BibitemShut {NoStop}%
\bibitem [{\citenamefont {Stanzione}\ \emph {et~al.}(2017)\citenamefont
  {Stanzione}, \citenamefont {Barth}, \citenamefont {Gaffney}, \citenamefont
  {Gaither}, \citenamefont {Hempel}, \citenamefont {Minyard}, \citenamefont
  {Mehringer}, \citenamefont {Wernert}, \citenamefont {Tufo}, \citenamefont
  {Panda},\ and\ \citenamefont {Teller}}]{10.1145/3093338.3093385}%
  \BibitemOpen
  \bibfield  {author} {\bibinfo {author} {\bibfnamefont {D.}~\bibnamefont
  {Stanzione}}, \bibinfo {author} {\bibfnamefont {B.}~\bibnamefont {Barth}},
  \bibinfo {author} {\bibfnamefont {N.}~\bibnamefont {Gaffney}}, \bibinfo
  {author} {\bibfnamefont {K.}~\bibnamefont {Gaither}}, \bibinfo {author}
  {\bibfnamefont {C.}~\bibnamefont {Hempel}}, \bibinfo {author} {\bibfnamefont
  {T.}~\bibnamefont {Minyard}}, \bibinfo {author} {\bibfnamefont
  {S.}~\bibnamefont {Mehringer}}, \bibinfo {author} {\bibfnamefont
  {E.}~\bibnamefont {Wernert}}, \bibinfo {author} {\bibfnamefont
  {H.}~\bibnamefont {Tufo}}, \bibinfo {author} {\bibfnamefont {D.}~\bibnamefont
  {Panda}},\ and\ \bibinfo {author} {\bibfnamefont {P.}~\bibnamefont
  {Teller}},\ }in\ \href {https://doi.org/10.1145/3093338.3093385} {\emph
  {\bibinfo {booktitle} {Proceedings of the Practice and Experience in Advanced
  Research Computing 2017 on Sustainability, Success and Impact}}},\ \bibinfo
  {series and number} {PEARC17}\ (\bibinfo  {publisher} {Association for
  Computing Machinery},\ \bibinfo {address} {New York, NY, USA},\ \bibinfo
  {year} {2017})\BibitemShut {NoStop}%
\end{thebibliography}%

\end{document}